\documentclass[mreferee]{gji}
\usepackage{timet}
\usepackage[centertags]{amsmath}
\usepackage{graphicx}
\usepackage[round, colon]{natbib}
\usepackage[multidot]{grffile}
\usepackage{hyperref}
\usepackage[utf8]{inputenc}
\usepackage[pagewise]{lineno}

\hypersetup{
     colorlinks=true,         
     linkcolor=blue,          
     citecolor=blue,          
     urlcolor=blue,           
}
\title[Microtremor H/V(z, f) modeling in marine environment]
  {A generalized  theory for full microtremor horizontal-to-vertical $[H/V(z, f)]$ spectral ratio interpretation in offshore and onshore environments}
\author[A. M. Lontsi et al.]
        {\\Agostiny Marrios Lontsi$^1$\thanks{Corresponding author:
        Agostiny Marrios Lontsi, agostiny.lontsi@sed.ethz.ch}, Antonio García-Jerez$^{2,8}$,
        Juan Camilo Molina-Villegas$^{3,9}$,
        Francisco José S\'anchez-Sesma$^4$,\\
        Christian Molkenthin$^5$,
Matthias Ohrnberger$^6$, Frank Kr\"uger$^6$,
        Rongjiang Wang$^7$, Donat F\"ah$^1$\\
 $^1$ Swiss Seismological Service, ETH Z\"urich, Switzerland,\\
  $^2$ Departamento de Qu\'imica y F\'isica,  Universidad de Almeria, Espa\~na,\\
$^3$ Facultad de Ingenier\'ias, Universidad de Medell\'in, Colombia,\\
  $^4$ Instituto de Ingenier\'ia, Universidad Nacional Aut\'onoma
  de M\'exico,  M\'exico,\\
   $^5$ Institute of Mathematics, University
 of Potsdam, Germany,\\
   $^6$ Institute of earth and environmental Science, University
 of Potsdam, Germany,\\
   $^7$ GFZ German Research Centre for Geosciences, Potsdam, Germany\\
   $^8$ Instituto Andaluz de Geofísica, Universidad de Granada, Espa\~na,\\
   $^9$ Departamento de ingeniería civil, Facultad de Minas, Universidad Nacional de Colombia – Sede Medellín, Colombia.
}
\date{\today}
\pagerange{\pageref{firstpage}--\pageref{lastpage}}
\volume{XXX}
\pubyear{2018}

\begin{document}
\label{firstpage}

\maketitle

\section*{Abstract}
\addcontentsline{toc}{section}{Abstract}

Advances in the field of seismic interferometry have provided a basic theoretical interpretation to the 
full spectrum of the microtremor horizontal-to-vertical spectral ratio $[H/V(f)]$. The interpretation has 
been applied to ambient seismic noise data recorded both at the surface and 
at depth. The new algorithm, based on the diffuse wavefield assumption, has been used in 
inversion schemes to estimate seismic wave velocity profiles that are useful 
input information for engineering and exploration seismology both for earthquake 
hazard estimation and to characterize surficial sediments. However, until now, 
the developed algorithms are only suitable for on land environments with no 
offshore consideration. Here, the microtremor $H/V(z, f)$ modeling is extended for applications 
to marine sedimentary environments for a 1D layered medium. The layer propagator matrix 
formulation is used for the computation of the required Green’s functions. 
Therefore, in the presence of a water layer on top, the propagator matrix for the uppermost layer 
is defined to account for the properties of the water column. As an 
application example we analyze eight simple canonical layered earth models. Frequencies ranging 
from $0.2$ to $50$ Hz are considered as they cover a broad wavelength interval and aid in practice to 
investigate subsurface structures in the depth range from a few meters to a 
few hundreds of meters. Results show a marginal variation of $8$ percent at most for the fundamental 
frequency when a water layer is present. The water layer leads to variations in $H/V$ peak amplitude of up to  
$50$ percent atop the solid layers.

\section{Introduction}

Over  the past decades, using the single-station microtremor horizontal-to-vertical ($H/V$) spectral 
ratio  as a method for shallow subsurface characterization has  
attracted a number of site investigation studies both on 
land (e.g. \citealp{bib:bard1998,bib:faehetal2003,bib:scherbaumetal2003,bib:lontsietal2015-hv,bib:lontsietal2016-dc-hv,bib:garcia-jerezetal2016,bib:pina-floresetal2017,bib:spicaetal2018,bib:garciajerezetal2019}) and  in marine environment 
(e.g. \citealp{bib:huerta-lopezetal2003,bib:muyzert2007,bib:overduinetal2015}). 
The interest in the method is mainly due to 
its practicability, its cost efficiency, and 
the minimum investment  effort 
during microtremor (ambient noise or passive seismic) survey campaigns. 
The generic engineering parameter directly estimated from 
the spectrum of the microtremor $H/V$ spectral ratio 
is the site fundamental 
frequency (e.g. \citealp{bib:nakamura1989,bib:lachetandbard1994}). 
The  fundamental frequency of a site generally corresponds to the frequency for 
which the microtremor $H/V$ spectral ratio reaches its maximum amplitude. 

Although the peak frequency is relatively well understood, 
this is not straightforward for secondary peaks as they
could represent higher modes or materialize the presence of 
more than one strong 
contrast in the  subsurface lithology. It is therefore important 
in the analysis 
to use a physical formulation for the $H/V$ spectral ratio that not only 
accounts for the full spectrum (including first and  subsequent 
secondary peaks), but also includes 
all  wave constituent parts. 
Based upon the advances in seismic noise 
interferometry (e.g. \citealp{bib:lobkisandweaver2001,bib:shapiroandcampillo2004,bib:curtisetal2006,bib:wapenaarandfokkema2006,bib:sens-schoenfelderandwegler2006,bib:gouedardetal2008b}), \citet{bib:sanchezetal2011b}
 proposed a physical model for the interpretation of the full spectrum of the 
microtremor $H/V$ spectral ratio. This has been extended to include 
receivers at depths \citep{bib:lontsietal2015-hv}. This additional 
information from receivers at depth is an added value during the 
velocity imaging process \citep{bib:lontsietal2015-hv,bib:lontsi2016,bib:spicaetal2018}.
As the interpretation effort focuses on  
the  $H/V$ spectral ratio acquisition on land, 
no significant effort has been made for the marine 
acquisition counterpart. An early study for a station on the seafloor was 
performed by \citet{bib:huerta-lopezetal2003}, assuming that the 
wavefield is due to the 
propagation of an incident plane SH body wave.
With the evolving technology in borehole 
acquisition seismic instruments and data 
transmission (e.g. \citealp{bib:stephenetal1994}),
there is a growing need for efficient subsea exploration and  
 geohazard estimation as reported by  \citet{bib:djikpesseetal2013}. 

Here we further extend the diffuse field model \citep{bib:sanchezetal2011b,bib:lontsietal2015-hv} to 
allow 
for the interpretation of the 
$H/V(z, f)$ both in
marine sedimentary environment and on land even though applicability 
to marine environments is emphasized.  

The Thomson-Haskell propagator  matrix \citep{bib:thomson1950,bib:haskell1953}
is  used to relate the displacement and stress for SH and P-SV waves 
at two points within an 
elastic 1D layered medium.
The use of the propagator matrix formulation  
allows us to easily include a propagator for a layer on top that accounts
 for the 
properties of the water layer and to subsequently compute the Green's 
function for points at different depths.
The classical Thomson-Haskell method  is unstable when  
 waves become evanescent. To remedy this issue, many attempts have 
been made (e.g \citealp{bib:knopoff1965,bib:dunkin1965,bib:abo-zano1979,bib:kennettandkerry1979,bib:harvey1981,bib:wang1999}). 
Here, we use the orthonormalization approach by \citet{bib:wang1999} which 
preserves the original Thomson-Haskell matrix algorithm and avoid the 
loss of precision by inserting an additional procedure that 
makes \textit{in-situ} base vectors orthonormal.

A synthetic analysis is performed on eight simple canonical earth 
models. 
The models differ by the presence of soft sediment structures with different
overall thickness (two in total) and the  presence of a water 
column with varying depth at the top.  
The first sediment structure is a very simple one layer over a 
half-space earth model 
and  the second is a realistic structural model obtained from site characterization at Baar, a municipality in the Canton of Zug, Switzerland.
The $H/V$ spectral ratio is estimated for 
frequencies ranging from $0.2$ to $50$ Hz. The effects of the water column 
on the $H/V$ 
spectrum at selected depths are interpreted. 

\section{Microtremor $H/V$ spectral ratio: A physical interpretation}
Here, the main steps linking the microtremor $H/V(z, f)$ spectral ratio 
to the elastodynamic Green's functions are presented. The basic expressions 
for SH and P-SV wave contributions to the Green's functions and some 
considerations for numerical integration  are summarized. 

\subsection{$H/V(z, f)$ interpretation: Onshore case}

Starting from a three-component ambient vibration data, the 
microtremor $H/V$ spectral 
ratio at a given point at the earth surface or at 
depth (onshore: Figure \ref{fig:layered-media} without water layer) for a known frequency $f$ is estimated 
using Equation~\ref{eq:hv-def}.

\begin{eqnarray}
	\begin{aligned}
	 H/V(z, f) &= \sqrt{\frac{E_{1}(z, f)+ E_{2}(z, f)}
                        {E_{3}(z, f)}},
	\end{aligned}
	\label{eq:hv-def}
\end{eqnarray}
 \noindent where
$  {E_{m}}(z, f) = \rho \omega^2 \langle {u_{m}}(z, f){u_{m}}^{*}(z, f)\rangle$ is 
physically regarded as the directional energy density, 
$\rho$ is the mass density, $\omega$ is the angular frequency
 and $u_m$ ($m=1,2,3$) is the  recorded displacement wavefield in 
the orthogonal direction $m$. The indexes $m=1, 2$ correspond to the horizontal 
components while $m=3$ corresponds to the index for the vertical component. The summation convention for repeated indexes is not applied here.
The symbol $*$ stands for complex conjugate.
 Using interferometric principles under the diffuse field assumption,
 it can be shown that the average of the autocorrelation of 
the displacement field is proportional to the imaginary part of the Green's function 
assuming the source and the receivers are at the same  point
 (\citealp{bib:sanchezsesmaetal2008,bib:sniederetal2009}, 
see a summary in Appendix 
\ref{appen:green}).
Equation~\ref{eq:hv-def} in terms of the Green's function is expressed as: 

\begin{eqnarray}
	\begin{aligned}
         H/V(z, f)  &= \sqrt{\frac{\text{Im}[G_{11}(z, z, f)]+\text{Im}[G_{22}(z, z, f)]
                }{\text{Im}[G_{33}(z, z, f)]}}
		=\sqrt{\frac{2\text{Im}[G_{11}(z, z, f)]
                }{\text{Im}[G_{33}(z, z, f)]}}
	\end{aligned}
	\label{eq:hv-depth}
\end{eqnarray}

We are therefore left with the computation of the  Green's functions $G_{11}=G_{22}$ and $G_{33}$.
The elastodynamic Green's function 
in a 1D elastic layered  medium (onshore: Figure \ref{fig:layered-media} without water layer)
is the set of responses for unit harmonic loads in the 
three directions. Using cylindrical coordinates the contribution of the 
radial-vertical (P-SV) and transverse (SH) motions are decoupled. 
Therefore, it suffices to solve each case separately using 
the integration on the horizontal wavenumber \citep{bib:bouchonandaki1977}.

\subsection{SH and P-SV contribution to the Green's function}

\begin{figure}
   \centering
    \includegraphics[width=\linewidth]{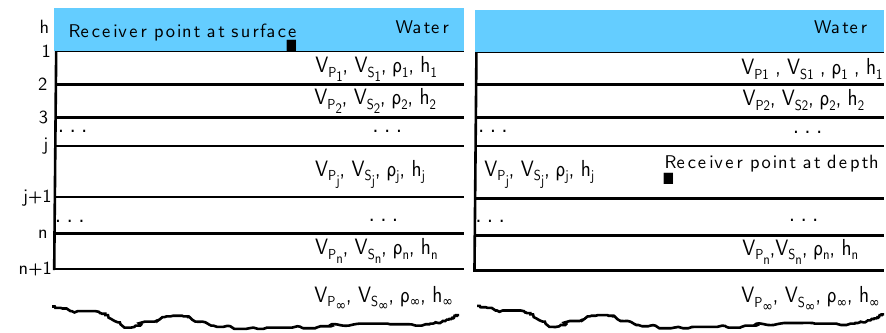}
   \protect\caption{Schematic representation of a 1D layered medium. The representation without the 
	water layer on top corresponds to the onshore case and the representation with water 
	layer corresponds to the offshore case. For the representation on the left, the 
	receiver location is at the earth surface when no water layer is present (onshore) and at the 
	water (lake, sea, ocean) bottom when the water layer is present (offshore). 
	For the representation on the right, the receiver location is at 
	depth. Except for the water layer in the offshore case 
	where the shear wave velocity is zero, any other layer $j$ either onshore or 
	offshore is 
	characterized by the seismic parameters $V_{\text{P}_{j}}$, $V_{\text{S}_{j}}$,
	$\rho_j$, $h_j$, $Q_{\text{P}_{j}}$, and $Q_{\text{S}_{j}}$.
}                                                                               
\label{fig:layered-media}
\end{figure}

Assuming the subsurface structure can be approximated by a stack of homogeneous 
layers over a half-space as depicted in Figure \ref{fig:layered-media} where for example the $j^{th}$ layer 
is characterized in the onshore case by the compressional  
wave velocity $V_{\text{P}_{j}}$ 
, the shear wave velocity $V_{\text{S}_{j}}$, the density $\rho_j$, the layer thickness $h_j$, and the 
attenuation parameters $Q_{\text{P}_{j}}$ and $Q_{\text{S}_{j}}$ for the P- and and S-wave 
respectively, Im$[G_{11}]$, Im$[G_{22}]$ 
and Im$[G_{33}]$ are given by:

\begin{eqnarray}
	\begin{aligned}
	\text{Im}[G_{11}] = \frac{1}{4\pi}\int_{0}^{\infty} \text{Im} \left[g_{11\text{SH}}\right]kdk
	+ \frac{1}{4\pi}\int_{0}^{\infty} \text{Im} \left[g_{11\text{PSV}}\right]kdk 
	\end{aligned}
	\label{eq:img11-img22}
\end{eqnarray}

\begin{eqnarray}
    \begin{aligned}
		\text{Im}[G_{33}] = \frac{1}{2\pi}\int_{0}^{\infty} \text{Im}\left[g_{33\text{PSV}}\right]kdk\cdot 
    \end{aligned}
	\label{eq:img33}
\end{eqnarray}

Because of symmetry, $\text{Im}[G_{11}] =\text{Im}[G_{22}]$.
Here $k$ is the radial wavenumber. 
The kernels   $g_{11\text{SH}}$, $g_{11\text{PSV}}$ and  
$g_{33\text{PSV}}$ correspond to the SH and P-SV wave 
contributions. The explicit dependence of  
 $g_{11\text{SH}}$, $g_{11\text{PSV}}$,   
and  $g_{33\text{PSV}}$ on the Thomson-Haskell propagator matrix (2x2 for SH waves and 4x4 for P-SV waves) for the  
layered elastic earth model presented in Figure \ref{fig:layered-media}
 are given in  Appendices \ref{appen:sh} and \ref{appen:p-sv}.

\subsection{$H/V(z, f)$ interpretation: Offshore case}

In the particular case where the top layer is a 
perfect homogeneous   water layer, the shear wave velocity and  
shear modulus do not exist (are null).  Substituting directly the 
corresponding  properties into the formulae of the 4x4   
propagator matrix for P-SV waves 
(Equations \ref{eq:psv-propagatorbegin}-\ref{eq:psv-propagatorend}) leads to a 
singular matrix. In this limiting case, there is an alternative approach to
consider  $P$-waves along the water column. 
A pseudo 4x4 propagator  matrix  $\mathbf{P}_{\text{pseudo}}$ 
is defined (Equation \ref{eq:pseudo}) and treated  
as in the onshore case \citep{bib:herrmann2008}. 

\begin{eqnarray}
        \begin{aligned}
		\mathbf{P}_{\text{pseudo}} &= \begin{pmatrix} 1 & 0 & 0 & 0\\
			0 & \cosh(\gamma h) & 0 & -\dfrac{\gamma}{\rho\omega}\sinh(\gamma h)\\
          0 & 0 & 1 & 0 \\
		0 & -\dfrac{\rho\omega^2}{\gamma} \sinh(\gamma h) & 0 & \cosh(\gamma h)\end{pmatrix} \\
        \end{aligned},
	\label{eq:pseudo}
\end{eqnarray}
\noindent where $\gamma=\sqrt{k^2-\omega^2/V_{\text{P}}^2}$ represents the vertical wavenumber 
for P-wave in water and $h$ the thickness of the  
water column.  
 A full derivation of $\mathbf{P}_{\text{pseudo}}$ is presented in  Appendix \ref{sect:pseudo}.

\subsection{Considerations for numerical implementation}

For the numerical integration, equations \ref{eq:img11-img22} and  \ref{eq:img33} 
 are transformed into a summation assuming virtual sources spread  along the
horizontal plane with  generic spacing $L$ \citep{bib:bouchonandaki1977}. 
The parameter $L$ also defines the integration step
$dk = \dfrac{2\pi}{L} $. The vertical wavenumbers  $\gamma_j$ and 
$\nu_j$ for, respectively, $P-$ and $S-$waves in the $j^{th}$ layer  relate 
to the horizontal wavenumbers $k$ by:

    \begin{eqnarray}
        \begin{aligned}
		\gamma_{j} = \sqrt{k^{2} - \frac{\omega^2}{V_{\text{P}_{j}}^2} }
        \end{aligned}
        \label{eq:gamma}
    \end{eqnarray}
\begin{eqnarray}
        \begin{aligned}
		\nu_{j} = \sqrt{k^{2} - \frac{\omega^2}{V_{\text{S}_{j}}^2} }.
        \end{aligned}
        \label{eq:nu}
\end{eqnarray}

Because of pole singularities of the kernel that account for 
the effects of surface waves,  
 a stable integration on the real axis can be performed if  
a correction term $\omega_I$ is added to the frequency to shift 
the poles of the kernel from the real axis, so that the effective 
frequency is:
\begin{eqnarray}
\omega = 2\pi f + \omega_I i, 
\end{eqnarray}
where $i$ is the unit imaginary number. $\omega_I$ is chosen as the smallest constant 
that effectively smooth out the kernels.
Anelastic attenuation of P- and S-wave energy
is considered by defining complex seismic wave 
velocities (See e.g. \citealp{bib:mueller1985}).

Additional considerations are made to avoid the loss-of-precision  
associated with the Thomson-Haskell propagator 
matrix when waves become evanescent. A
 numerical procedure  is inserted into the matrix propagation loop to make all 
determined displacement vectors
\textit{in-situ} orthonormal \citep{bib:wang1999}. The orthonormalization  
procedure, as implemented here, for both  surface downward- and infinity upward wave 
propagation of the determined base vectors 
is presented in Appendix 
\ref{appen:orthonormalization}.

\section{Synthetic analysis using  canonical and realistic earth models}

For testing the presented algorithm, the directional energy 
density profile for a homogeneous half-space is computed.  
Table \ref{tab:0LOH-model} presents the model parameters for this simple earth structure defined as a Poisson solid.
\begin{table}
        \protect\caption[Seismic parameters for a homogeneous.]{Seismic parameters for a homogeneous half-space.
        The model is used to estimate the directional energy density profile
        with normalized depth.
}
\label{tab:0LOH-model}
 \centering
 \begin{tabular}{|l| l| l | l|l|l|}
 \hline
 $h$ (m)     & $V_P$ (m/s) & $V_S$ (m/s) & $\rho \left( \text{kg/m}^{3} \right)$ & $Q_P$ & $Q_S$ \\
 \hline
 $\infty$   & 1732 & 1000  & 2000  & 100 & 100 \\
\hline
 \end{tabular}
 \end{table}
\begin{figure}
        \centering
        \includegraphics[width=0.9\linewidth]{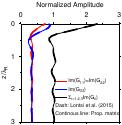}
	\protect\caption{Normalized energy density 
	profiles (Im($G_{11}$), Im($G_{22}$), Im($G_{33})$, and the total directional energy density) for the three orthogonal directions
	estimated using (1) the algorithm  
	based on the propagator matrix formulation (thin continuous line) and  (2) the algorithm 
	based on the global matrix formulation for a 
	layered medium (dashed thick line; \citealt{bib:lontsietal2015-hv}). 
	The depth is normalized with 
the Rayleigh wavelength. There is a good agreement between the two approaches for 
	Green's functions estimation. 
	Input parameters used in the modeling are defined  in Table~\ref{tab:0LOH-model}.
}
\label{fig:DED-profile}
        \end{figure}
The energy variation with depth as depicted in Figure \ref{fig:DED-profile}
 shows a good
agreement with the known theory regarding the energy partition for a diffuse
wavefield (e.g. \citealp{bib:weaver1985,bib:pertonetal2009}). For depths larger
than approximately 1.5 times the Rayleigh  wavelength, there is almost no
surface wave energy contribution and the energy is equal
for the three orthogonal directions (Figure \ref{fig:DED-profile}).

Further tests are performed by considering a simple one layer over a half-space (1LOH) and 
a realistic subsurface structure.
 The realistic earth model has been obtained from site characterization 
at Baar, Canton Zug, Switzerland \citep{bib:hobigeretal2016}. 
The parameters for the simple one layer  over 
a half-space model together with those of the realistic earth structure
  used in the second  test are 
 presented in Table~\ref{tab:1LOH-model}. 
The 1LOH structural  model represents a
 very simple soft-soil characterized by 
a  constant shear wave
velocity ($V_S$) of $200$ m/s, a velocity contrast of $5$ in $V_S$ and an 
overall sediment cover of $25$ m. The realistic earth model at Baar 
has velocity contrast in $V_S$ of about $4$  between the 
 sediment layer overlaying the half-space and the half-space. Here, the overall sediment cover 
is about  $100$ m.   
In comparison to $V_S$ values that 
remain almost constant, water saturated sediment offshore have compressional 
wave velocities  estimates that are much larger than the onshore 
 values. 

\begin{figure*}
        \centering
\includegraphics[width=\linewidth]{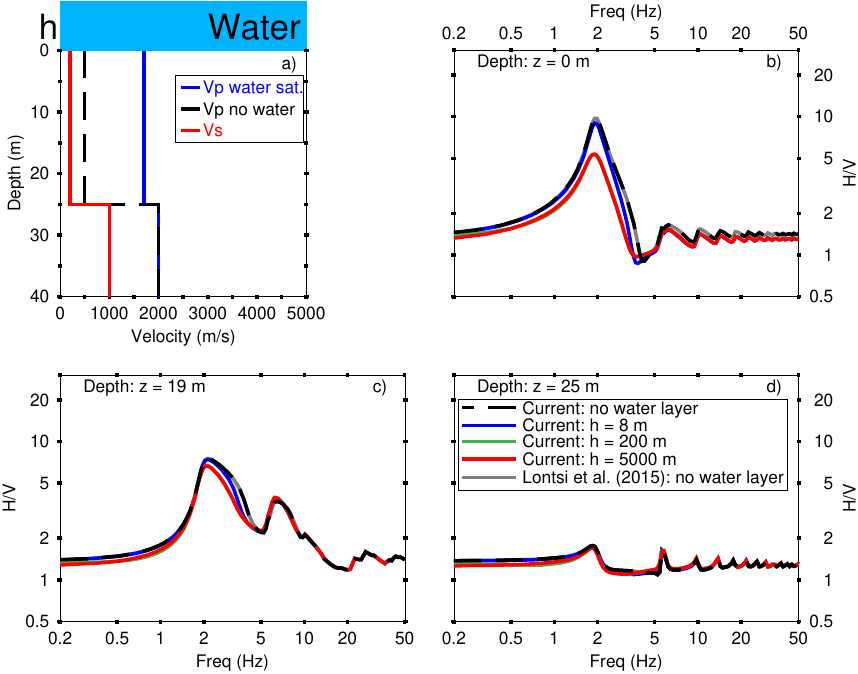}
 \protect\caption[a) Seismic parameters for
a very simple soft soil layer over a half-space.]{
a) Seismic parameters for
	a simple soft soil layer over a half-space (defined in Table~\ref{tab:1LOH-model}). 
	The P-wave velocity in water is set to $1500$ m/s. The water-saturated sediments have the 
	velocity set to $1700$ m/s (see solid blue profile). The shear-wave velocity ($V_S$) 
	profile is set unchanged in the presence of the water layer.  
	b) Comparison between $H/V$ spectral ratios at the solid-liquid interface ($z=0$ m). 
c) Comparison between $H/V$ spectral ratios at $19$ m depth and d) at $25$ m depth. 
	The gray curve is obtained using the extended global matrix formulation for receivers at depth when no water layer is present (see \citealt{bib:lontsietal2015-hv}). The 
	computed $H/V$ for 
	a synthetic water layer of $200$ and $5000$ m shows nearly the same results and 
	are almost overlayed with each other; see green and red curves.
}
\label{fig:compare-methods-M2.1}
        \end{figure*}

\begin{figure*}
        \centering
\includegraphics[width=\linewidth]{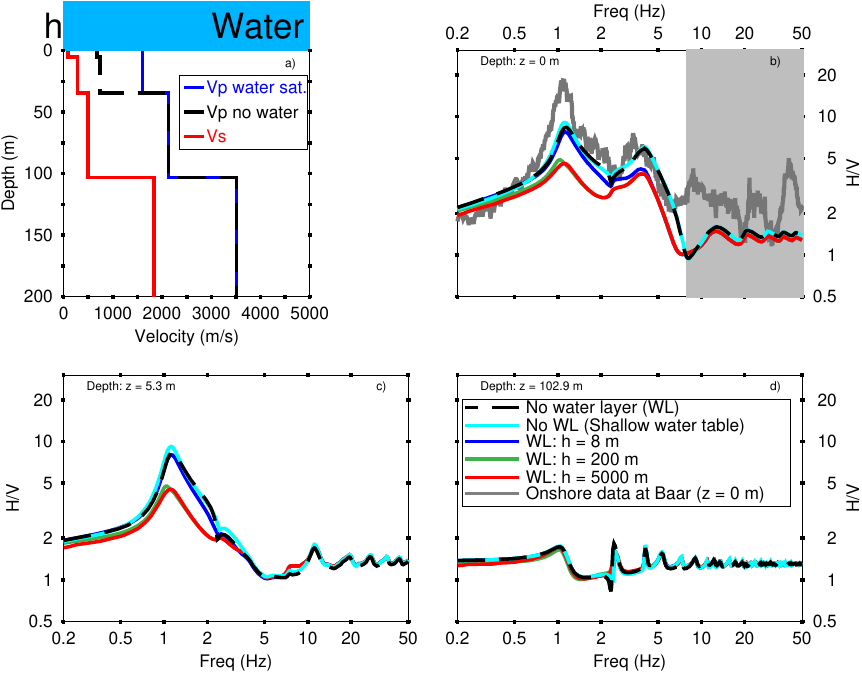}
 \protect\caption{
a) Seismic parameters for
 a realistic earth model (defined in Table~\ref{tab:1LOH-model}. The 
	P-wave velocity for water  is set to $1500$ m/s. 
The P-wave velocity for water-saturated sediments are represented with the 
	solid blue profile. The shear-wave velocity ($V_S$) profile is set 
unchanged in the presence of the water layer. In addition, a sedimentary environment
 with a very shallow water table is considered. The $V_P$ for 
the sediment   in this onshore case was set to $1600$ m/s. 
b) Comparison between $H/V$ spectral ratios at the solid-free surface and  
solid-liquid interface ($z = 0$ m). For the solid-free surface interface (onshore), 
field data exist and are used for validation of the presented algorithm (solid gray curve). 
Frequencies above $8$ Hz (light gray box) were not used for the profile estimation.
c) Comparison between $H/V$ spectral ratios at $5.3$ m within the sediment column 
	and d) at $102.9$ m depth (sediment bedrock interface).
}
\label{fig:compare-methods-SBAS}
        \end{figure*}

Figure \ref{fig:compare-methods-M2.1}a, and respectively Figure \ref{fig:compare-methods-SBAS}a
present the seismic velocity profiles ($V_P$ and $V_S$) for 
the two investigated structural models. Considered $V_P$ profiles
 for the water saturated sediments are represented by the 
blue solid line. The 
corresponding $H/V(z, f)$ spectral ratio 
 without a water layer (onshore) and with water layer (offshore)  are plotted together 
 for different depths. This representation allows   for a visual appraisal of the effect of the water column 
 (Figures \ref{fig:compare-methods-M2.1}~b-d and
\ref{fig:compare-methods-SBAS}~b-d). 
\begin{table}
 \protect\caption{Test models consisting of one and three solid layers
	over a half-space (onshore). Offshore cases, characterized by $V_S = 0$ m/s 
	are built by considering a water layer on top.
	In the case where the water layer is considered, the $V_P$ velocities for the sediment 
	at the bottom of the water column are modified to 
	account for the saturation with water. Considered values for $V_P$ are shown in parenthesis 
	in the appropriated column.  Scenarios for different water environments ranging from 
	shallow to deep are considered. 
	The $H/V$ spectral ratios at three different locations (surface + two additional depths) 
	for these two illustrative cases
 are presented in Figures \ref{fig:compare-methods-M2.1}~and~\ref{fig:compare-methods-SBAS}.
}
\label{tab:1LOH-model}
 \centering
 \begin{tabular}{|l| l| l | l|l|l|}
 \hline
\multicolumn{6}{l}{One-layer over a half-space}\\
 \hline
 $h$ (m)     & $V_P$ (m/s) & $V_S$ (m/s) & $\rho \left( \text{kg/m}^{3} \right)$ & $Q_P$ & $Q_S$ \\
 \hline
\textcolor{blue}{8$^a$}(\textcolor{cyan}{200$^b$},\textcolor{red}{5000$^c$})   & \textcolor{blue}{1500} & \textcolor{blue}{0}  & \textcolor{blue}{1000}  & \textcolor{blue}{99999} & \textcolor{blue}{99999} \\
 25   & 500 (\textcolor{blue}{1700}) & 200  & 1900  & 100 & 100 \\
 $\infty$  & 2000 & 1000  & 2500 & 200 & 200 \\
 \hline
\multicolumn{6}{l}{Realistic earth model at Baar, Canton Zug}\\
 \hline
         \textcolor{blue}{8$^a$}(\textcolor{cyan}{200$^b$},\textcolor{red}{5000$^c$})   & \textcolor{blue}{1500} & \textcolor{blue}{0}  & \textcolor{blue}{1000}  & \textcolor{blue}{99999} & \textcolor{blue}{99999} \\
         5.3 & 672.8 (\textcolor{blue}{1600})  & 85.6 & 2000 & 100& 100\\
         29.2 & 738.9 (\textcolor{blue}{1600}) & 284.3 & 2000& 100 &100\\
         68.4& 2135.6 & 500.0 & 2000& 100 &100\\
         $\infty$ & 3512.2 & 1841.1 & 2300& 100 &100\\
 \hline
 \end{tabular}

        $^a$ Thin water layer. $^b$ Lake environment. $^c$ Deep ocean environment.
 \end{table}

The $H/V(z, f)$ spectral ratio computed with the propagator matrix algorithm for the 1LOH are 
calibrated with results obtained using the global matrix formulation 
 approach as presented by \citealp{bib:lontsietal2015-hv} for receivers at depth
 when no water layer is present (compare solid gray and 
 dashed black dashed lines on Figures \ref{fig:compare-methods-M2.1}~b-d).  
The two approaches (propagator matrix and global matrix formulations) 
  provide $H/V$ spectral ratios that agree with each other for all tested 
receivers locations for the onshore case.

The presented algorithm is further  
 used to assess the variations of  the $H/V$ spectral ratios,
at the surface and at depth, due to the presence of the water layer. 
To this end, the structural models presented in Table~\ref{tab:1LOH-model} with 
three different water-layer thicknesses  ($8$, $200$, $5000$ m) are used. The water layer thicknesses are 
selected to reflect different water environments ranging from shallow lake to deep sea.
For the one layer-over-halspace (1LOH) structural model and for a scenario of shallow 
water environment with $8$ m water 
column, we observe at frequencies above $2$ Hz (peak frequency) an 
amplitude variation. Further scenarios with moderate ($200$ m) to 
deep ($5000$ m) water layer indicate  that the  amplitude variations extend to 
low frequencies and reach up to  $50\%$ around 
the $H/V$ spectral ratio peak amplitude for the receiver at the surface.
Only marginal relative 
variations are observed for the H/V peak frequency when the water layer is present.
The amplitude variation as well as the marginal peak frequency variation
  observed for the one layer-over-halspace in different water environments
are also  valid for the  
realistic earth model at Baar. For this particular test site, we further consider that the water 
table is very shallow and investigate the onshore  scenario  with water saturated sediments cover. 
The $V_P$ velocity for the first two layers was set to $V_P=1600$ m/s to consider the saturation with water. The 
resulting H/V spectral ratio computed  at different depths indicates that 
changes in Vp do have influence on the shape of the H/V spectral 
ratio in the frequency band ranging from about 1 to 3 Hz for receivers at 
the surface and at depths, although very 
minor (see Figure~\ref{fig:compare-methods-SBAS}b-d). 
 At Baar, onshore ambient vibrations data from array recordings are available. 
The surface waves analysis allowed to extract the average  seismic velocity profiles of the underlying 
subsurface structure (for more details, see \citealt{bib:hobigeretal2016}).
 The estimated velocity profiles did not 
account for $H/V$ information beyond $8$ Hz shown in the light gray box (Figure~\ref{fig:compare-methods-SBAS}b).
The $H/V$ spectral ratio from the array central station is 
used for calibration (gray curve in Figure~\ref{fig:compare-methods-SBAS}b).

Within the sediment column, and 
for all considered water column thicknesses, the variability of the $H/V$ spectral ratio 
is observed up to a certain cut-off frequency. This cut-off frequency is about  $5$ Hz at $19$ m depth 
for the  1LOH structural model. For the realistic earth model at Baar and for a receiver at about $5.3$ m depth, 
the cut-off frequency is about  $10$ Hz. 
In the last case where the receiver is located at the bedrock interface, a marginal $H/V$ amplitude 
variations are observed (Figures~\ref{fig:compare-methods-M2.1}d and \ref{fig:compare-methods-SBAS}d). For both models, the low-frequency peak 
corresponds well with the fundamental resonance of SH waves in the structure. 
In the case of the layer above the half-space it is given by the simple 
relationship $f_{0} = \dfrac{V_{S}}{4H}$, where $H$ is the thickness of 
the sediment column and $V_{S}$ is the shear wave 
velocity  ($Vs = 200$ m/s and $H = 25$ m, Figure~\ref{fig:compare-methods-M2.1}a).
 For the realistic model at Baar (Figure~\ref{fig:compare-methods-SBAS}a),
 the peak frequency can be estimated by using the simple 
expression found by \citet{bib:tuanetal2016} with about $10\%$ deviation.
  Secondary peaks for the simple one 
 layer over a 
 half-space (Figure~\ref{fig:compare-methods-M2.1}a) satisfy the 
 relationship $f_{n} = \dfrac{V_{S}}{4H}(2n+1)$. For the realistic earth 
 model at Baar (Figure~\ref{fig:compare-methods-SBAS}), the second dominant peak at about 
 $4$ Hz corresponds to the response of the top layer characterized by  a shear wave velocity $Vs = 85$ m/s. 
The corresponding 
 impedance contrast is about $3.34$. A weak impedance contrast of about $1.76$ exists between the 
 second and third layer. Additional peaks (Gray box Figure~\ref{fig:compare-methods-SBAS} b) would depend on very shallow features 
not represented by  the considered velocity  model.  

\section{Understanding the $H/V$ amplitude variation}

\begin{figure*}
        \centering
\includegraphics[width=\linewidth]{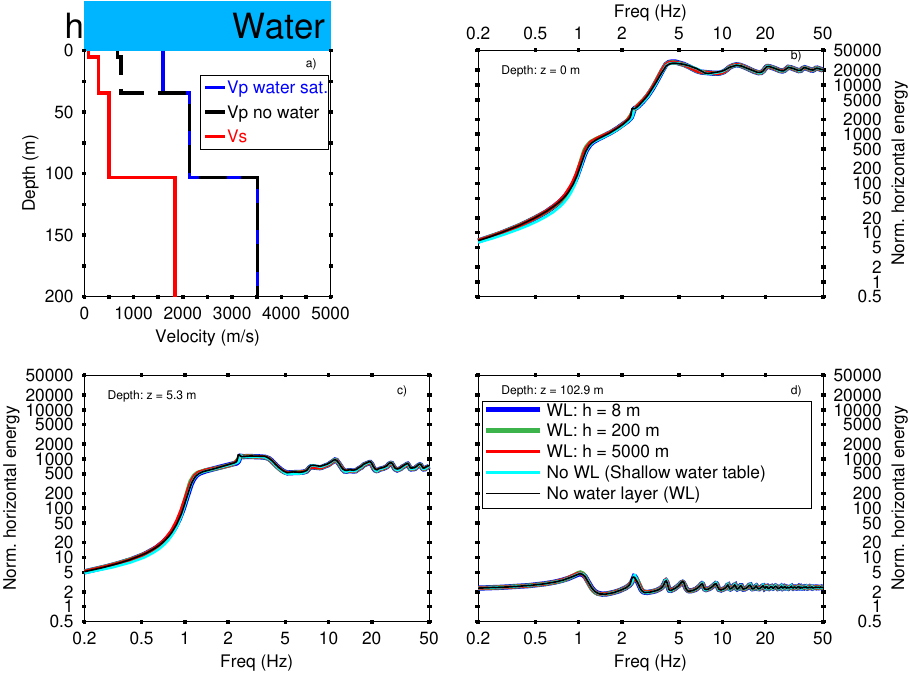}
 \protect\caption{Horizontal directional energy density variation at the seabottom using the structural earth model at Baar.
Different water layer thicknesses are considered.
}
\label{fig:energy-horizontal-SBAS}
        \end{figure*}

\begin{figure*}
        \centering
\includegraphics[width=\linewidth]{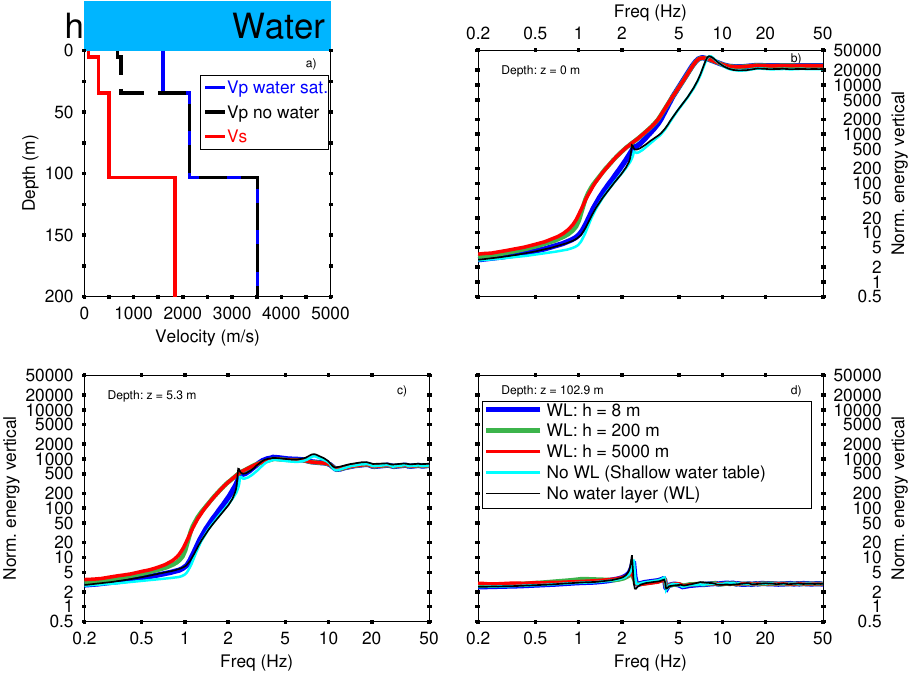}
 \protect\caption{Vertical directional energy density variation at the seabottom using the structural earth model at Baar. 
Different water layer thicknesses are considered. 
}
\label{fig:energy-vertical-SBAS}
        \end{figure*}

The observed amplitude variation of the H/V in the presence of the water 
layer are investigated by analyzing the modeled directional  
energy densities (DED) both  on  the  
horizontal and vertical components. The earth model at Baar is used 
for the analysis. Figures~\ref{fig:energy-horizontal-SBAS} 
and \ref{fig:energy-vertical-SBAS}  show the modeled DED for the 
horizontal and vertical components respectively. Considered scenarios include an 
earth model (1) without water layer, (2) with no water layer but very shallow 
water table, (3) with  water layer with 8-, 200-, and 5000 m. 
   It comes out that the energy 
on the horizontal component is not sensitive to the presence of the water layer. This 
is understood as no shear wave is expected to  propagate in the considered
 ideal fluid (no viscosity). On the contrary, we observe significant energy 
variations on the vertical component that can be associated with multiple 
energy reverberations in the water layer.  

\begin{figure*}
        \centering
\includegraphics[width=\linewidth]{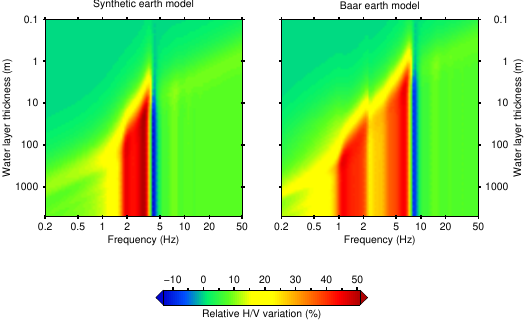}
 \protect\caption{Relative variation of the H/V spectral ratio in the presence of the 
water layer  with respect the the H/V spectral ratio when no water layer is present. 
Water layer thicknesses ranging between 0.1 and 5000 m, sampled on 
a logarithmic scale are considered. Left: One layer over half-space structure. Right: Realistic earth model at Baar.}
\label{fig:water-depth}
        \end{figure*}

We further assess the dependence of the amplitude variation with a much larger number of 
water layer thicknesses scenario. For this purpose, the relative variation of H/V spectral ratio when there is water layer is  studied. Figure~\ref{fig:water-depth} depicts this relative variations in percent for a wide range of water columns.
It can be observed that the presence of a shallow water layer mainly has effects on the very 
high frequency. Deep water environment affect the amplitude of the H/V spectral ratio on a very broad frequency range. 

\section{Conclusions}
A theoretical model based on the diffuse field approximation is proposed
 for the estimation of the horizontal-to-vertical ($H/V(z, f)$) spectral ratio 
on land and in marine environment. The propagator matrix (PM) method has been 
used to compute the Green's function in a 1D layered medium 
including a liquid layer atop. For onshore cases, the modeled $H/V(z, f)$ spectral curves 
are compared with estimations from the  
  global matrix (GM) approach and show good agreement for the 
considered synthetic  structural models. 
 In comparison to the GM method,
the PM  provides an efficient approach for modeling the $H/V$ spectral ratio
within marine environments. 
Modeling results indicate that the 
$H/V$ spectral ratio is sensitive to the presence of a water layer overlying
subseabed sediments. 
$H/V$ relative amplitude variations are observed in the complete 
 considered frequency range 
($0.2 - 50$ Hz) for  deep water environment and  may reach approximately $50\%$ 
around the peak frequency. 
The amplitude decrease in the $H/V$ peak 
can be understood as large P-wave energy on the vertical component 
result from multiple reverberation in the water column. The $H/V$ data available at Baar (onshore) are 
used to validate the presented algorithm for a receiver at the surface. For computed cases, 
changes in the fundamental 
 frequency are marginal. In addition, primary resonances  occur at frequencies 
that satisfy the relationships used in practical applications. Secondary resonances in the  
1LOH corresponds to overtones while in the realistic case at Baar, they materialize the 
response of the  subsequent layers within the sediment column.

\section*{Acknowledgments}

Our thanks to F. Luzón for his comments and suggestions 
and to J. E. Plata and G. Sánchez of the USI-Inst. Eng.-UNAM for locating useful 
references. 
This work has been partially supported, through the sinergia program, by 
the Swiss National Science Foundation (grant 171017), by the Spanish Ministry of 
Economy and Competitiveness (grant CGL2014-59908), by the European Union with ERDF, by DGAPA-UNAM (Project IN100917), and by the 
Deutsche Forschungsgemeinschaft (DFG) through grant CRC 1294 "Data Assimilation" (Project B04). 
Constructive comments from the editors Jörg Renner 
and Jean Virieux, Hiroshi Kawase and two anonymous reviewers helped to improve the quality of the manuscript.

\appendix

\section{Ambient noise - Green's function - representation theorem - 
	cross-correlation - Directional Energy  Density - Equipartition}
\label{appen:green}

Seismic sources at the origin of the ambient noise 
wavefield are ubiquitous and may be  
at surface or/and at depth. Generated seismic waves are back-scattered in the 
subsurface.  The recorded noise 
wavefield at a seismic station after a lapse time  large enough  
compared to the mean travel time of ballistic 
waves (e.g direct waves, first reflected waves) contains information regarding the 
underlying subsurface structure. 

Let us assume that there is an asymptotic regime with a stable supply 
of energy that constitutes the background illumination. 
This condition, for what it shares with the radiative transport, 
is called diffuse field. In an unbounded elastic medium 
a harmonic diffuse field is considered random, isotropic and 
equipartitioned.
The stabilization of the S to P energy ratio is reached 
asymptotically for long lapse times \citep{bib:pauletal2005}. 
 Within such a field
the Directional Energy Density (DED), $E_{m}$, for a given 
orthogonal direction $m$ is given in terms of the averaged autocorrelation of the 
 displacement wavefield and it is proportional to the imaginary part of the 
Green's function of  the system for the source and the receiver at the same location.
This is expressed in Equation \ref{eq:auto-corr-energ}:  
\begin{eqnarray}
\begin{aligned}
  E_{m}(\mathbf{x},f) =\rho \omega^2 \langle u_{m}(\mathbf{x},f)u_{m}^{*}(\mathbf{x},f)\rangle \propto \text{Im}[G_{mm}(\mathbf{x},\mathbf{x},f)]
\label{eq:auto-corr-energ}
\end{aligned}
\end{eqnarray}
Here $\rho = \rho(\mathbf{x})$ is the mass density at point $\mathbf{x}$, and $\omega=2\pi f$ is the circular frequency.  No summation 
over the repeated index $m$ is assumed.
In practice, the DED is estimated from the autocorrelation (power spectrum) of the recorded 
ambient noise wavefield and averaged over short time windows.
This is equivalent to average over directions if the field is isotropic. 
For subsurface imaging purposes, the DED is computed from the imaginary part of 
the Green's function (Equation~\ref{eq:auto-corr-energ}). 

In order to demonstrate the validity of Equation~\ref{eq:auto-corr-energ} and  
establish the  proportionality factor, 
a simple homogeneous, isotropic, elastic medium is considered. 
Therefore, the analytical expression to the Green's 
function is known \citep{bib:sanchezsesmaetal2008}.

The displacement field $u_{i}(\mathbf{x},\omega)$ produced by a body force 
$f_i$ 
at a given point $\mathbf{x}$ of an elastic solid is described by the Newton's law of displacements (Equation \ref{eq:navier}).

\begin{eqnarray}
\dfrac{\partial}{\partial x_j} \left ( c_{ijkl}\dfrac{\partial u_{l}(\mathbf{x},\omega)}{\partial x_k} \right )
	+ \omega^2\rho u_{i}(\mathbf{x},\omega) 
	= - f_{i}(\mathbf{x},\omega)
\label{eq:navier}
\end{eqnarray}   

Equation \ref{eq:navier} is often called elastic wave equation or Navier equation. 
Here $c_{ijkl}$ is the stiffness tensor. The Einstein summation convention 
is assumed, i.e repeated index implies summation over the range of that index. 

From Equation \ref{eq:navier}, it is possible to derive the classical Somigliana representation theorem (e.g. \citealp{bib:wapenaarandfokkema2006,bib:vanmanenetal2006,%
        bib:sniederetal2007,bib:sanchezsesmaetal2008,bib:sanchez-sesmaetal2018}):

\begin{eqnarray}
	\begin{aligned}
		&u_{m}(\mathbf{x_A},\omega)=\\
		&\int_\varGamma \left [ G_{im}(\mathbf{x},\mathbf{x_A},\omega)t_i(\mathbf{x},\omega) 
        - T_{im}(\mathbf{x},\mathbf{x_A},\omega)u_i(\mathbf{x},\omega) \right ] d\varGamma_x
        + \int_{V} f_{i}(\mathbf{x},\omega)G_{im}(\mathbf{x_A},\mathbf{x},\omega)dV_x
		\end{aligned}
        \label{eq:representation-theo1}
\end{eqnarray}
in which one has the displacement field for $\mathbf{x_A}$ being a point at $V$ inside 
the surface $\varGamma$ in terms of body forces and the boundary values of displacements
and tractions.
Here $G_{im}(\mathbf{x_A},\mathbf{x},\omega)$ and  $T_{im}(\mathbf{x_A},\mathbf{x},\omega)$ are the Green's functions for displacements and tractions when the harmonic unit force are in the direction $m$. $f_i(\mathbf{x})$ is the body force distribution.   
 $t_{i}(\mathbf{x},\omega)$ and $T_{im}(\mathbf{x_A},\mathbf{x},\omega)$
are defined by Equation \ref{eq:tiandTim}.
\begin{eqnarray}
\begin{aligned}
	&t_i(\mathbf{x},\omega) 
	= n_j(\mathbf{x}) \left ( c_{ijkl}\dfrac{\partial u_{l}(\mathbf{x},\omega)}{\partial x_k} \right )\\
	&T_{im}(\mathbf{x_A},\mathbf{x},\omega) = n_j(\mathbf{x}) \left ( c_{ijkl}\dfrac{\partial G_{lm}(\mathbf{x}, \mathbf{x_A},\omega)}{\partial x_k} \right )	\cdot
\label{eq:tiandTim}
\end{aligned}
\end{eqnarray}

By considering for the internal point $\mathbf{x_B}$ an harmonic 
body force $f_i(\mathbf{x})\equiv\delta(\mathbf{x}-\mathbf{x_B})\delta_{in}$ in 
the direction $n$  and setting for the field the time-reversed solution, 
then $u_i(\mathbf{x})\equiv G_{in}(\mathbf{x},\mathbf{x_B},\omega)$,
$t_i(\mathbf{x})\equiv T_{in}(\mathbf{x},\mathbf{x_B},\omega)$, 
and Equation \ref{eq:representation-theo1} becomes:
\begin{eqnarray}
\begin{aligned}
	&\int_\varGamma \left [ T_{im}(\mathbf{x},\mathbf{x_A},\omega)G_{in}^*(\mathbf{x},\mathbf{x_B},\omega)
	- T_{in}^*(\mathbf{x},\mathbf{x_B},\omega)G_{im}(\mathbf{x},\mathbf{x_A},\omega) \right ] d\varGamma_x = \\ 
	&-G_{mn}^*(\mathbf{x_A},\mathbf{x_B},\omega) + G_{mn}(\mathbf{x_A},\mathbf{x_B},\omega)
\end{aligned}
\end{eqnarray}
which is re-written changing $\mathbf{x}$ by $\mathbf{\xi}$,
 to represent boundary points on $\varGamma$, as: 
\begin{eqnarray}
	\begin{aligned}
		& 2\mathbf{i}G_{mn}(\mathbf{x_A},\mathbf{x_B},\omega) =\\
		&-\int_\varGamma \left [ G_{mi}(\mathbf{x_A},\mathbf{\xi},\omega)T_{in}^*(\mathbf{\xi},\mathbf{x_B},\omega)
        - G_{ni}^*(\mathbf{x_B},\mathbf{\xi},\omega)T_{im}(\mathbf{\xi},\mathbf{x_A},\omega) \right ] d\varGamma_{\xi} \cdot
\end{aligned}
\label{eq:corr-reptheo}
\end{eqnarray}
The Equation \ref{eq:corr-reptheo}
 is a correlation-type representation theorem. 
A similar form has been presented by \citet{bib:vanmanenetal2006}. 
Then because the theorem in Equation \ref{eq:corr-reptheo} is valid for any surface $\varGamma$, 
it follows that if the field is diffuse at the envelope,i.e. the net flux of energy is null, it is also 
diffuse at any point within the heterogeneous medium   

Starting from the analytical expressions for $G_{im}$ and $T_{im}$ in the farfield
(\citealp{bib:sanchezsesmaandcampillo2006,bib:sanchezsesmaetal2008};
see, e.g., \citealp{bib:dominguezandabascal1984} for the full expression 
of $G_{im}$ and $T_{im}$), 
it can be  demonstrated, that for random and uncorrelated sources, the resulting illumination, 
after a lapse time large enough compared to the travel time of 
ballistic waves, is an equipartitioned diffuse field.
Therefore, the right hand side of 
equation~\ref{eq:corr-reptheo}  is proportional to the azimuthal 
average of crosscorrelation of the displacement field.
\begin{eqnarray}
\begin{aligned}
  \text{Im}[G_{mn}(\mathbf{x_A},\mathbf{x_B})]=-(2\pi\xi_S)^{-1}k^3 \langle u_{m}(\mathbf{x_A})u_{n}^{*}(\mathbf{x_B})\rangle 
\end{aligned}
\end{eqnarray}

Where $k$ is the shear wave number and $\xi_S$ is the  average energy density
of shear waves and represents a measure of the strength of the diffuse
illumination. 
 Assuming the source 
and the receiver are at the same location ($\mathbf{x_A}=\mathbf{x_B}=\mathbf{x}$), 
one can thus write \citep{bib:sanchezsesmaetal2008}:
\begin{eqnarray}
\begin{aligned}
  \text{Im}[G_{mm}(\mathbf{x},\mathbf{x})]=-(2\pi\xi_S)^{-1}k^3 \langle | u_{m}(\mathbf{x})|^2\rangle 
\end{aligned}
\end{eqnarray}

An alternative approach linking the azimuthal average of 
cross-correlation to the imaginary part of the Green's function under diffuse 
assumption in the farfield, and without prior knowledge of the full analytical 
expression of the Green's function was presented by
  \citet{bib:sniederetal2009}.

\section{Estimating the SH waves contribution to the 
imaginary part of the Green's function}
\label{appen:sh}
\subsection{Receiver at the surface}

Following  \citealp{bib:akiandrichards2002}, and for the displacement $u_2=v$, 
the SH-wave equation in an arbitrary layer $j$ presented in 
Figure \ref{fig:layered-media} is given in linear elasticity by:

\begin{eqnarray}
\dfrac{\partial^2 v}{\partial t^2} = 
\dfrac{\mu_j}{\rho_j}\left ( \dfrac{\partial^2 v}{\partial x^2} + \dfrac{\partial^2 v}{\partial z^2} \right )
\label{eq:wave2dsh}
\end{eqnarray}
A solution to the Equation \ref{eq:wave2dsh} can be of the form:

\begin{eqnarray}
v = l_1(z,w,k)\exp[ i(kx -\omega t)]
\label{eq:wave2dsh-sol}
\end{eqnarray}
and the associated shear stresses:
\begin{eqnarray}
\begin{aligned}
	\tau_{yz}&=\mu_j \dfrac{\partial l_1}{\partial z}\exp [i(kx -\omega t)] \\
			&= l_2 \exp[ i(kx -\omega t)]\\
      \tau_{xy} &= ik\mu_j l_1 \exp[ i(kx -\omega t)]\cdot  
\end{aligned}
\label{eq:stressfield}
\end{eqnarray}
From Equation \ref{eq:stressfield}, the differential 
Equation \ref{eq:eqdiff1} is obtained.

\begin{eqnarray}
\dfrac{d l_1}{d z} = \dfrac{1}{\mu_j }l_2 
\label{eq:eqdiff1}
\end{eqnarray}

From Newton's second law, one gets: 

\begin{eqnarray}
\dfrac{\partial \tau_{xy}}{\partial x} + \dfrac{\partial \tau_{yz}}{\partial z} = 
\rho_j \dfrac{\partial^2 v}{\partial t^2}
\label{eq:wave2dsh-N}
\end{eqnarray}

This leads to 

\begin{eqnarray}
\dfrac{d l_2}{d z} = (k^2\mu_j - \omega^2 \rho_j) l_1 
\label{eq:eqdiff2}
\end{eqnarray}

Equations \ref{eq:eqdiff1} and \ref{eq:eqdiff2} lead to the 
system of first order differential equations~\ref{eq:eqdiff}.

\begin{eqnarray}
\dfrac{d}{d z} \dbinom{l_1}{l_2} = \begin{pmatrix} 0&\dfrac{1}{\mu_j}\\
k^2\mu_j - \omega^2\rho_j& 0\\\end{pmatrix} \dbinom{l_1}{l_2} 
\label{eq:eqdiff}
\end{eqnarray}

This equation  is of the form:

\begin{eqnarray}
\dfrac{d \mathbf{l}}{d z} = \mathbf{A}_j \mathbf{l} 
\label{eq:eqdiffmatrix}
\end{eqnarray}
where 
\begin{eqnarray}
 \mathbf{l} = \dbinom{l_1}{l_2} 
\end{eqnarray}
and
\begin{eqnarray}
 \mathbf{A}_j= \begin{pmatrix} 0&\dfrac{1}{\mu_j}\\
k^2\mu_j - \omega^2\rho_j& 0\end{pmatrix}\cdot
\end{eqnarray}

Assuming a homogeneous layer medium  (i.e., the layers properties are constant), 
the solution to Equation \ref{eq:eqdiffmatrix} within the $j^{th}$ layer defined by $z_j$ and $z_{j+1}$ is  
given by:
\begin{eqnarray}
\dbinom{l_1}{l_2}_{z_{j+1}} = \mathbf{P}_j \dbinom{l_1}{l_2}_{z_j} 
\end{eqnarray}
where
\begin{eqnarray}
	\mathbf{P}_j= \exp[\mathbf{A}_j(z_{j+1} - z_j)] 
\end{eqnarray}
Using linear algebra properties, it can be shown that if 
 the matrix $\mathbf{A}$ is diagonalizable, then there exists an invertible $\mathbf{L}$ so that 

\begin{eqnarray}
	\mathbf{A}= \mathbf{LeL}^{-1} 
	\label{eq:linear-algebrabegin}
\end{eqnarray}
where $\mathbf{L}$ is the matrix of eigenvectors of $\mathbf{A}$; $\mathbf{L}^{-1}$ its inverse,  and  $\mathbf{e}$ 
is the eigenvalue matrix.
The series expansion of $\exp[\mathbf{A}(z_{j+1} - z_j)]$ using  
\begin{eqnarray}
	\exp(\mathbf{A})=\sum_{k=0}^{\infty}\dfrac{\mathbf{A}^k}{k!} = \mathbf{I} + \mathbf{A} + \dfrac{\mathbf{A}^2}{2!} + \dfrac{\mathbf{A}^3}{3!} + \cdots  
\end{eqnarray}

allows to write:

\begin{eqnarray}
	\exp(\mathbf{L}^{-1}\mathbf{AL}) = \exp(\mathbf{e}) = \mathbf{E} =  \mathbf{L}^{-1}\left (\mathbf{I} + \mathbf{A} + \dfrac{\mathbf{A}^2}{2!} + \dfrac{\mathbf{A}^3}{3!} + \ldots \right )\mathbf{L} 
	=\mathbf{L}^{-1}[\exp(\mathbf{A})]\mathbf{L} 
	\label{eq:linear-algebraend}
\end{eqnarray} 

Where $\mathbf{E} = \exp(\mathbf{e})$.
For the problem investigated, $\mathbf{A}$ is 
replaced by $\mathbf{A}_j(z_{j+1} - z_j)$. 

The eigenvalues for the 2x2 $\mathbf{A}(z_{j+1} - z_j)$ for the SH wave propagation
 are obtained  by finding the roots of the 
second order polynomial defined by:

\begin{eqnarray}
	\det[\mathbf{A}_j(z_{j+1} - z_j) - \lambda I] = 0
\end{eqnarray}

This leads to 

\begin{eqnarray}
\lambda_1 = (z_{j+1} - z_j)\nu_j\\
\lambda_2 = -(z_{j+1} - z_j)\nu_j
\end{eqnarray}

where $\nu_j = \sqrt{k^2 - \dfrac{\omega^2\rho_j}{\mu_j}} = \sqrt{k^2 - \dfrac{\omega^2}{V_{S_j}^2}} $

The eigenvectors are obtained by solving the equations (for the two eigenvalues):

\begin{eqnarray}
	[\mathbf{A}_j(z_{j+1} - z_j) - \lambda_1 I]\begin{pmatrix} x_1 \\ x_2 \end{pmatrix} = 0
\end{eqnarray}

and 

\begin{eqnarray}
	[\mathbf{A}_j(z_{j+1} - z_j) - \lambda_2 I]\begin{pmatrix} x_1 \\ x_2 \end{pmatrix} = 0
\end{eqnarray}

Sample eigenvectors are therefore: \\

for $\lambda_1$:

\begin{eqnarray}
(x_1, x_2) = (1,\mu_j\nu_j)x_1 
\end{eqnarray}

and for $\lambda_2$:

\begin{eqnarray}
(x_1, x_2) = (1,-\mu_j\nu_j)x_1 \cdot
\end{eqnarray}

The eigenvectors can be arranged in the matrix

\begin{eqnarray}
	\mathbf{L}_j = \begin{pmatrix} 1 & 1\\
\mu_j\nu_j & -\mu_j\nu_j \end{pmatrix}
\end{eqnarray}

and the inverse of $\mathbf{L}_j$ is given by:

\begin{eqnarray}
	\mathbf{L}_j^{-1} = \begin{pmatrix} \dfrac{1}{2} & \dfrac{1}{2\mu_j\nu_j}\\
\dfrac{1}{2} & \dfrac{-1}{2\mu_j\nu_j} \end{pmatrix}
\end{eqnarray}

The matrix $\mathbf{E}_j$ is given by:

\begin{eqnarray}
	\mathbf{E}_j = \begin{pmatrix} \exp[\nu_j (z_{j+1} - z_j)] & 0\\
0 &  \exp[-\nu_j (z_{j+1} - z_j)] \end{pmatrix}
\end{eqnarray}

The eigenvalue problem has also been studied by  \citealp{bib:gantmacher1959,bib:gilbertandbackus1966}.

The propagator (or layer) matrix can therefore be written as:

\begin{eqnarray}
	\begin{aligned}
		\mathbf{P}_j  &= \mathbf{L}_j\mathbf{E}_j\mathbf{L}_j^{-1} \\&= \begin{pmatrix} 1 & 1\\
\mu_j\nu_j & -\mu_j\nu_j \end{pmatrix}\begin{pmatrix} \exp[\nu_j (z_{j+1} - z_j)] & 0\\
0 &  \exp[-\nu_j (z_{j+1} - z_j)] \end{pmatrix}\begin{pmatrix} \dfrac{1}{2} & \dfrac{1}{2\mu_j\nu_j}\\
\dfrac{1}{2} & \dfrac{-1}{2\mu_j\nu_j} \end{pmatrix} 
\label{eq:eigenp}
	\end{aligned}
\end{eqnarray}

This operation leads to 

\begin{eqnarray}
	\mathbf{P}_j = \begin{pmatrix} \cosh[\nu_j(z_{j+1}-z_j)]&\dfrac{1}{\mu_j\nu_j}\sinh[\nu_j(z_{j+1}-z_j)]\\
\mu_j\nu_j \sinh[\nu_j(z_{j+1}-z_j)] & \cosh[\nu_j(z_{j+1}-z_j)] \end{pmatrix}
\end{eqnarray}

In the next steps, the propagator matrix as defined by Equation \ref{eq:eigenp} is 
 used. This representation allows to 
introduce a manipulation matrix to avoid the instability problem in the high 
frequency range.

For a $n$-layer over half-space system, we obtain:

\begin{eqnarray}
\dbinom{l_1}{l_2}_{z_{n+1}} = \mathbf{P}_n\mathbf{P}_{n-1}...\mathbf{P}_1 \dbinom{l_1}{l_2}_{z_1} 
\label{eq:eigenproduct}
\end{eqnarray}

By introducing Equation \ref{eq:wave2dsh-sol} into Equation \ref{eq:wave2dsh-N}, a second 
order differential equation is obtained for $l_1$ where the solution can be written for layer 1 
in the form:

\begin{eqnarray}
l_1 = \acute{S}_1 \exp(\nu_1 z) + \grave{S}_1 \exp(-\nu_1 z) 
\label{eq:l1}
\end{eqnarray}
where $\acute{S}_1$ and $\grave{S}_1$ are constant representing the amplitude 
of upgoing- and downgoing SH waves.\\
Equations \ref{eq:eqdiff1} and  \ref{eq:l1}  lead to 

\begin{eqnarray}
l_2 = \mu_1\nu_1\acute{S}_1 \exp(\nu_1 z) - \mu_1\nu_1\grave{S}_1 \exp(-\nu_1 z) 
\label{eq:l2}
\end{eqnarray}

Equation \ref{eq:l1} and \ref{eq:l2} combine to
\begin{eqnarray}
\renewcommand\arraystretch{1.3}
 \dbinom{l_1}{l_2} = \begin{pmatrix} 
\exp(\nu_1 z) & \exp(-\nu_1 z)\\
\mu_1\nu_1\exp(\nu_1 z) & -\mu_1\nu_1\exp(-\nu_1 z)
 \end{pmatrix}\dbinom{\acute{S}_1}{\grave{S}_1}
\end{eqnarray}

Without loss of generality, we have for the half-space:
\begin{eqnarray}
\begin{aligned}
 \dbinom{l_1}{l_2}_{n+1} &= \begin{pmatrix} 
\exp(\nu_{n+1} z) & \exp(-\nu_{n+1} z)\\
\mu_{n+1}\nu_{n+1}\exp(\nu_{n+1} z) & -\mu_{n+1}\nu_{n+1}\exp(-\nu_{n+1} z)
 \end{pmatrix}\dbinom{\acute{S}_{n+1}}{\grave{S}_{n+1}}\\
		&= \begin{pmatrix} 1 & 1\\
\mu_{n+1}\nu_{n+1} & -\mu_{n+1}\nu_{n+1} \end{pmatrix}\dbinom{\acute{S}_1\exp(\nu_{n+1}z)}{\grave{S}_1\exp(-\nu_{n+1}z)}\\
		&=\mathbf{L}_{n+1}\dbinom{\acute{S}_{n+1}\exp(\nu_{n+1}z)}{\grave{S}_{n+1}\exp(-\nu_{n+1}z)}\\
		&=\mathbf{L}_{n+1}\begin{pmatrix} \exp(\nu_{n+1} z_{n+1} ) & 0\\
0 &  \exp(-\nu_{n+1}z_{n+1}) \end{pmatrix} \dbinom{\acute{S}_{n+1}}{\grave{S}_{n+1}}
\end{aligned}
\end{eqnarray}

In the half-space, there is no up-going waves, therefore $\acute{S}_{n+1}=0$, so that:
\begin{eqnarray}
\begin{aligned}
	\dbinom{l_1}{l_2}_{n+1} &= \mathbf{L}_{n+1}\begin{pmatrix} \exp(\nu_{n+1} z_{n+1}) & 0\\
0 &  \exp(-\nu_{n+1}z_{n+1}) \end{pmatrix} \dbinom{0}{\grave{S}_{n+1}}\\
	&=\mathbf{L}_{n+1}\begin{pmatrix} \exp(\nu_{n+1} z_{n+1} ) & 0\\
0 &  \exp(-\nu_{n+1}z_{n+1}) \end{pmatrix}\dbinom{0}{1}\grave{S}_{n+1}
	\label{eq:displacementatdepth}
\end{aligned}
\end{eqnarray}
 
Using Equation \ref{eq:eigenp}, \ref{eq:eigenproduct}, and \ref{eq:displacementatdepth}, we have: 

\begin{eqnarray}
	\mathbf{P}_n\mathbf{P}_{n-1}...\mathbf{P}_1 \dbinom{l_1}{l_2}_{1} = \mathbf{L}_{n+1}\begin{pmatrix} \exp(\nu_{n+1} z_{n+1} ) & 0\\
0 & \exp(-\nu_{n+1}z_{n+1}) \end{pmatrix}\dbinom{0}{1}\grave{S}_{n+1} 
\end{eqnarray}

This leads, for the displacement at the surface to:
\begin{eqnarray}
 \dbinom{l_1}{l_2}_{1} &= \mathbf{P}_1^{-1}...\mathbf{P}_{n-1}^{-1}\mathbf{P}_n^{-1}\mathbf{L}_{n+1}\dbinom{0}{1}\grave{S}_{n+1} \exp(-\nu_{n+1}z_{n+1})
	\label{eq:extractg22}
\end{eqnarray}

Where $\mathbf{P}_{n}^{-1} =  \mathbf{L}_n\mathbf{E}_n^{-1}\mathbf{L}_n^{-1}$

Lets set $\mathbf{C}_{n+1}=\dbinom{0}{1}$ and $\mathbf{Y}_{n+1}=\mathbf{L}_{n+1}\mathbf{C}_{n+1}$

At the surface load point ($z=0$), $l_1=v=g_{22\text{SH}}$ (the integrand of 
interest) and $l_2=1$.

Equation  \ref{eq:extractg22} can be solved for $g_{22\text{SH}}$ at the surface. 

The Green's function in the 1D layered medium is obtained by integration over the horizontal 
wavenumber: 
\begin{eqnarray}
 \text{Im}\left[G_{22}^{SH}(z, f)\right] = \text{Im}\left[G_{11}^{SH}(z, f)\right] = \frac{1}{4\pi}\int_{0}^{\infty} 
 \text{Im}\left[g_{22\text{SH}}\right]kdk
\label{eq:imgsh}
\end{eqnarray}

Note that a correction factor $\dfrac{k}{4\pi}$ has been introduced in the kernel. 
This is trivial in cylindrical coordinates when the radius and azimuthal components 
are set to zero.

\subsection{Receiver at depth}

For a receiver at depth, the displacement-stress just under the load point which is assumed to be at the interface $j$ can be written as 
follows (compare Equation \ref{eq:extractg22}): 
\begin{eqnarray}
\begin{aligned}
 \dbinom{l_{b1}}{l_{b2}}_{z_j} = \mathbf{P}_j^{-1}...\mathbf{P}_{n-1}^{-1}\mathbf{P}_n^{-1}\mathbf{L}_{n+1}\dbinom{0}{1}\grave{S}_{n+1} \exp(-\nu_{n+1}z_{n+1})
    \label{eq:extractg22-depth1}
\end{aligned}
\end{eqnarray}

On the other hand, the result just above the source would be: 
\begin{eqnarray}
\begin{aligned}
\dbinom{l_{u1}}{l_{u2}}_{z_j} = \mathbf{P}_{j-1}\mathbf{P}_{j-2}...\mathbf{P}_{1} \dbinom{l_1}{l_2}_{z_1}
\label{eq:fromthesurface}
\end{aligned}
\end{eqnarray}.

The boundary conditions at the load point at depth are given (1) for the upper layer by
 $l_{u1}$ = $g_{22\text{SH}}$; and $l_{u2}$ = $\tau_{u}$ and (2) 
for the  bottom layers by: $l_{b1}$ = $g_{22\text{SH}}$ and $l_{b2}$ = $\tau_{b}$. 
The unit load at the source is defined such that 
$\tau_{b} - \tau_{u} = 1$

Equation \ref{eq:fromthesurface} can be rewritten as

\begin{eqnarray}
\begin{aligned}
\dbinom{l_{u1}}{l_{u2}}_{z_{j}}&=
\mathbf{P}_{j-1}\mathbf{P}_{j-2}...\mathbf{P}_{1} \dbinom{v_s}{0}_{z_1}\\
							&= \mathbf{P}_{j-1}\mathbf{P}_{j-2}...\mathbf{P}_{1} \dbinom{1}{0}v_s
\label{eq:fromthesurface2}
\end{aligned}
\end{eqnarray} 

Lets set $\mathbf{Y}_{1}=\dbinom{1}{0}$ as the basic displacement-stress solution at the surface. This basic vector is propagated downwards from the 
surface to the source. So that:

\begin{eqnarray}
\dbinom{Y_{u1}}{Y_{u2}}=
 \mathbf{P}_{j-1}\mathbf{P}_{j-2}...\mathbf{P}_{1} \dbinom{1}{0}\cdot
\end{eqnarray}

Respectively, the fundamental vector  of plane-wave amplitude $\dbinom{0}{1}$ at 
the half-space can be propagated upwards to the source:

\begin{eqnarray}
	\dbinom{Y_{b1}}{Y_{b2}}=\mathbf{P}_j^{-1}...\mathbf{P}_{n-1}^{-1}\mathbf{P}_n^{-1}\mathbf{L}_{n+1}\dbinom{0}{1}
\end{eqnarray}

The set of boundary conditions allows to extract $g_{22\text{SH}}$ as:

\begin{eqnarray}
g_{22\text{SH}}=\frac{Y_{u1}Y_{b1}}{Y_{u1}Y_{b2}-Y_{u2}Y_{b1}}
\label{eq:g22}
\end{eqnarray}

For this SH case, the Green's function in the 1D layered medium is given by:
\begin{eqnarray}
 \text{Im}\left[G_{22}^{SH}(z, f)\right] = \text{Im}\left[G_{11}^{SH}(z, f)\right] 
 = \frac{1}{4\pi}\int_{0}^{\infty} \text{Im}\left[g_{22\text{SH}}\right]kdk
\end{eqnarray}

The integral can be numerically computed by making, e.g.,  use of the discrete wavenumber approach.

\section{Computing the P-SV waves contribution to the
imaginary part of the Green's function}
\label{appen:p-sv}

Without loss of generality, consider the  inplane solution ( i.e. no dependence on 
the  $y$ coordinate)  
to the elastic  wave or Navier equation. The displacement-stress 
vector  $\mathbf{r}=(r_1, r_2, r_3, r_4)^T$ is obtained by the following expression
(see also e.g. \citealp{bib:akiandrichards2002}; Chap7, p263):

\begin{eqnarray}
\begin{aligned}
	u &= r_1(k,z,\omega)\exp [i(kx-\omega t)],\\
v &= 0,\\
	w &= ir_2(k,z,\omega)\exp [i(kx-\omega t)],
\end{aligned}
\end{eqnarray}
Here we used $(u_1, u_2, u_3) = (u, v, w)$.
Let set the stresses associated to displacements:

\begin{eqnarray}
\begin{aligned}
\tau_{zx} &= r_3(k,z,\omega)\exp [i(kx-\omega t)],\\
\tau_{zz} &= ir_4(k,z,\omega)\exp [i(kx-\omega t)].
\end{aligned}
\end{eqnarray}

Using  Hooke's and Newton's law for a homogeneous medium it can be shown that:

\begin{eqnarray}
\dfrac{d}{d z} \begin{pmatrix} r_1\\ r_2\\ r_3\\ r_4 \\\end{pmatrix}  
= \begin{pmatrix} 0& k & \dfrac{1}{\mu} & 0 \\
\dfrac{-k\lambda}{\lambda + 2\mu} & 0 & 0 & \dfrac{1}{\lambda + 2\mu}\\
\dfrac{4k^2\mu(\lambda + \mu)}{\lambda + 2\mu} - \omega^2\rho & 0 & 0 & \dfrac{k\lambda}{\lambda + 2\mu}  \\
0 & - \omega^2\rho & -k & 0 \end{pmatrix}
\begin{pmatrix} r_1\\ r_2\\ r_3\\ r_4 \\\end{pmatrix}
\label{eq:eqdiffmatrixpsv}
\end{eqnarray}
which is the first order differential equation for displacement-stress vector $\mathbf{r}$. $\lambda$ and $\nu$ are the Lamé parameters.

Assuming a layer homogeneous medium (i.e., the layers properties do not 
depend on the depth $z$ for a given layer),
the solution to Equation \ref{eq:eqdiffmatrixpsv} at two points $z_1$ and $z_2$ is
given by:
\begin{eqnarray}
\begin{pmatrix} r_1\\ r_2\\ r_3\\ r_4 \\\end{pmatrix}_{z_2} = \mathbf{P} \begin{pmatrix} r_1\\ r_2\\ r_3\\ r_4 \\\end{pmatrix}_{z_1} 
	\label{eq:psvsolution}
\end{eqnarray}
where
\begin{eqnarray}
	\mathbf{P}= \exp[\mathbf{A}(z_2 - z_1)] 
\end{eqnarray}

Where 

\begin{eqnarray}
	\mathbf{A} = \begin{pmatrix} 0& k & \dfrac{1}{\mu} & 0 \\
\dfrac{-k\lambda}{\lambda + 2\mu} & 0 & 0 & \dfrac{1}{\lambda + 2\mu}\\
\dfrac{4k^2\mu(\lambda + \mu)}{\lambda + 2\mu} - \omega^2\rho & 0 & 0 & \dfrac{k\lambda}{\lambda + 2\mu}  \\
0 & - \omega^2\rho & -k & 0 \end{pmatrix}
\end{eqnarray}

The eigenvalues for the 4x4 matrix $\mathbf{A}(z_2 - z_1)$ for the P-SV wave propagation 
are obtained  
by finding the roots of the
fourth order polynomial defined by:

\begin{eqnarray}
	\det[\mathbf{A}(z_2 - z_1) - a \mathbf{I}] = 0
\end{eqnarray}

This leads to:

\begin{eqnarray}
\begin{aligned}
a_1 &= \gamma = \sqrt{k^2 - \dfrac{\omega^2}{\alpha}}(z_2 - z_1)\\
a_2 &= \nu = \sqrt{k^2 - \dfrac{\omega^2}{\beta}}(z_2 - z_1)\\
a_3 &= -\gamma = -\sqrt{k^2 - \dfrac{\omega^2}{\alpha}}(z_2 - z_1)\\
a_4 &= -\nu = -\sqrt{k^2 - \dfrac{\omega^2}{\beta}}(z_2 - z_1)
\end{aligned}
\end{eqnarray}

$\alpha$ and $V_p$ and  $\beta$ and $V_s$ are used interchangeably.

Using linear algebra properties as presented in  
Equations~\ref{eq:linear-algebrabegin}-\ref{eq:linear-algebraend}, we obtain:

\begin{eqnarray}
	\mathbf{L} = \begin{pmatrix}
\alpha k & \beta\nu & \alpha k & \beta\nu \\
\alpha\gamma  & \beta k & -\alpha\gamma &  -\beta k\\
-2\alpha\mu k \gamma & -\beta\mu(k^2 + \nu^2) & 2\alpha\mu k \gamma & \beta\mu(k^2 + \nu^2) \\
-\alpha\mu(k^2 + \nu^2) & -2\beta\mu k \nu & -\alpha\mu(k^2 + \nu^2) & -2\beta\mu k \nu
\end{pmatrix}
	\label{eq:psv-propagatorbegin}
\end{eqnarray}

\begin{eqnarray}
	\mathbf{E}=\begin{pmatrix}
		\exp{[\gamma (z_2-z_1)]} & 0  & 0  & 0 \\
		0  & \exp{[\nu (z_2-z_1)]} &  0  & 0 \\ 
		0  & 0  & \exp{[-\gamma (z_2-z_1)]} &  0 \\
		0  & 0  & 0 & \exp{[-\nu (z_2-z_1)]}
\end{pmatrix}
\end{eqnarray}

\begin{eqnarray}
	\mathbf{L}^{-1}=\dfrac{\beta}{2\alpha\mu\gamma\nu\omega^2} 
\begin{pmatrix}
2\beta\mu k \gamma\nu & -\beta\mu\nu(k^2 + \nu^2) & -\beta k \nu & \beta\gamma\nu\\
-\alpha\mu\gamma(k^2 + \nu^2) & 2\alpha\mu k\gamma\nu & \alpha\gamma\nu & -\alpha k \gamma \\
2\beta\mu k\gamma\nu &  \beta\mu\nu(k^2 + \nu^2) & \beta k \nu & -\beta\gamma\nu\\
-\alpha\mu\gamma(k^2 + \nu^2) & -2\alpha\mu k\gamma\nu & -\alpha\gamma\nu & -\alpha k \gamma
\end{pmatrix} 
\label{eq:psv-propagatorend}
\end{eqnarray}

where $\mathbf{L}$, $\mathbf{E}$ are the corresponding eigenvector- and exponential of the 
eigenvalues matrices respectively. 

Equation~\ref{eq:psvsolution} can be rewritten as:

\begin{eqnarray}
\begin{aligned}
	\begin{pmatrix} r_1\\ r_2\\ r_3\\ r_4 \\\end{pmatrix}_{z_2} = \mathbf{P} \begin{pmatrix} r_1\\ r_2\\ r_3\\ r_4 \\\end{pmatrix}_{z_1} 
		= \mathbf{LEL}^{-1}\begin{pmatrix} r_1\\ r_2\\ r_3\\ r_4 \\\end{pmatrix}_{z_1}
\end{aligned}
\end{eqnarray}

Without loss of generality, for the elastic layer n with layer top labeled $n$, the displacement-stress
vector at the bottom interface labeled $n+1$ is given by

\begin{eqnarray}
\begin{aligned}
	\begin{pmatrix} r_1\\ r_2\\ r_3\\ r_4 \\\end{pmatrix}_{z_{n+1}} = \mathbf{P}_n \begin{pmatrix} r_1\\ r_2\\ r_3\\ r_4 \\\end{pmatrix}_{z_n} 
\end{aligned}
\end{eqnarray}

where 

\begin{eqnarray}
\begin{aligned}
	\mathbf{P}_n=\mathbf{L}_n\mathbf{E}_n\mathbf{L}_n^{-1}
\end{aligned}
\end{eqnarray}

For a $n$-layer over half-space earth model, we obtain:

\begin{eqnarray}
\begin{aligned}
	\begin{pmatrix} r_1\\ r_2\\ r_3\\ r_4 \\\end{pmatrix}_{z_{n+1}} = \mathbf{P}_n\mathbf{P}_{n-1}...\mathbf{P}_1 \begin{pmatrix} r_1\\ r_2\\ r_3\\ r_4 \\\end{pmatrix}_{z_1} 
\end{aligned}
\end{eqnarray}

It can also be shown that

\begin{eqnarray}
	\begin{aligned}
		&	\begin{pmatrix} r_1 \\ r_2 \\ r_3 \\ r_4 \end{pmatrix}_{n+1} = \\&\mathbf{L}_{n+1}
\begin{pmatrix}
\exp(\gamma_{n+1} z_{n+1}) & 0  & 0  & 0 \\
 0  & \exp(\nu_{n+1} z_{n+1}) &  0  & 0 \\ 
0  & 0  & \exp(-\gamma_{n+1} z_{n+1}) &  0 \\
 0  & 0  & 0 & \exp(-\nu_{n+1} z_{n+1})
\end{pmatrix}\begin{pmatrix} \acute{P}_{n+1} \\ \acute{S}_{n+1} \\ \grave{P}_{n+1} \\ \grave{S}_{n+1}\end{pmatrix}
	\end{aligned}
\end{eqnarray}

In the half-space, there is no up-going $P$ and $SV$ waves, therefore $\acute{P}_{n+1}=0$ and $\acute{S}_{n+1}=0$.

\begin{eqnarray}
\begin{aligned}
	&\begin{pmatrix} r_1 \\ r_2 \\ r_3 \\ r_4 \end{pmatrix}_{n+1} = \\&\mathbf{L}_{n+1}
\begin{pmatrix}
\exp(\gamma_{n+1} z_{n+1}) & 0  & 0  & 0 \\
 0  & \exp(\nu_{n+1} z_{n+1}) &  0  & 0 \\ 
0  & 0  & \exp(-\gamma_{n+1} z_{n+1}) &  0 \\
 0  & 0  & 0 & \exp(-\mu_{n+1} z_{n+1})
\end{pmatrix}\begin{pmatrix} 0 \\ 0 \\ \grave{P}_{n+1} \\ \grave{S}_{n+1}\end{pmatrix}\\
	&= \mathbf{L}_{n+1}\begin{pmatrix}
0 & 0  \\
 0  &  0 \\ 
1  & 0  \\
 0  & 1 \end{pmatrix}
\begin{pmatrix}\grave{P}_{n+1}\exp(-\gamma_{n+1} z_{n+1}) \\ \grave{S}_{n+1}\exp(-\mu_{n+1} z_{n+1})\end{pmatrix}
\end{aligned}
\end{eqnarray}

The later representation together with the defined manipulation matrix (Appendix \ref{appen:orthonormalization}) allow to 
propagate the orthonormal base vectors $(0, 0 , 1, 0)^T$ 
        and $(0, 0 , 0, 1)^T$, i.e., the 2x1 matrix in an efficient way and ultimately to avoid the loss of precision issue 
	associated with the Thomson-Haskell propagator matrix. See also \citet{bib:wang1999}.

\subsection{Receiver at the surface}
Harmonic horizontal load: for a receiver at the surface, the boundary conditions 
for a harmonic  horizontal load are the following:

\begin{eqnarray}
\begin{pmatrix} r_1 \\ r_2 \\ r_3 \\ r_4 \end{pmatrix}_{1}=
\begin{pmatrix} g_{11\text{PSV}} \\ g_{31\text{PSV}}/i \\ -1 \\ 0  \end{pmatrix}
\end{eqnarray}

Harmonic vertical load: for a receiver at the surface, the boundary conditions 
for a harmonic  vertical load are the following:

\begin{eqnarray}
\begin{pmatrix} r_1 \\ r_2 \\ r_3 \\ r_4 \end{pmatrix}_{1}=
\begin{pmatrix} g_{13\text{PSV}} \\ g_{33\text{PSV}}/i \\ 0 \\ -1/i  \end{pmatrix}
\end{eqnarray}

In the half-space, we have the following boundary conditions:
\begin{eqnarray}
\begin{pmatrix} \grave{P} \\ \grave{S} \\ \acute{P} \\ \acute{S}\end{pmatrix} = 
\begin{pmatrix} \grave{P} \\ \grave{S} \\ 0 \\ 0\end{pmatrix}\cdot 
\end{eqnarray}

For the harmonic horizontal load we then have:
\begin{eqnarray}
\begin{aligned}
\begin{pmatrix} \grave{P} \\ \grave{S} \\ 0 \\ 0\end{pmatrix} &=
	\mathbf{L_{n+1}}^{-1}\mathbf{P}_{n}\mathbf{P}_{n-1}...\mathbf{P}_{1}
\begin{pmatrix} g_{11\text{PSV}} \\ g_{31\text{PSV}}/i \\ -1 \\ 0  \end{pmatrix}
\end{aligned}
\end{eqnarray}

and for the harmonic vertical load:

\begin{eqnarray}
\begin{aligned}
\begin{pmatrix} \grave{P} \\ \grave{S} \\ 0 \\ 0\end{pmatrix} &=
	\mathbf{L_{n+1}}^{-1}\mathbf{P}_{n}\mathbf{P}_{n-1}...\mathbf{P}_{1}
\begin{pmatrix} g_{13\text{PSV}} \\ g_{33\text{PSV}}/i \\ 0 \\ -1/i  \end{pmatrix}\cdot
\end{aligned}
\end{eqnarray}

The two  equations above can be solved for $g_{11\text{PSV}}$ and $g_{33\text{PSV}}$.

The Green's function for the P-SV case in a 1D layered medium are then given by: 
\begin{eqnarray}
 \text{Im}\left[G_{22}^{\text{P-SV}}(z, f)\right] = \text{Im}\left[G_{11}^\text{{P-SV}}(z, f)\right] = \frac{1}{4\pi}\int_{0}^{\infty} \text{Im}\left[g_{11\text{PSV}}\right]kdk
\end{eqnarray}

\begin{eqnarray}
 \text{Im}\left[G_{33}^\text{{P-SV}}(z_F,f)\right] =  \frac{1}{2\pi}\int_{0}^{\infty} \text{Im}\left[g_{33\text{PSV}}\right]kdk
\end{eqnarray}

\subsection{Receiver at depth}

The displacement-stress vector from the half-space to the source/receiver 
can be written in terms of the amplitudes of the waves in the half-space as:

\begin{eqnarray}
\begin{aligned}
\mathbf{L}_{n+1}\begin{pmatrix} 0 & 0 \\ 0 & 0 \\ 1 & 0 \\ 0 & 1 \end{pmatrix}\begin{pmatrix} \grave{P} \\ \grave{S} \end{pmatrix} &=
\begin{pmatrix} r_1 \\ r_2 \\ r_3 \\ r_4 \end{pmatrix}_{n+1}=
	\mathbf{P}_{n}\mathbf{P}_{n-1}...\mathbf{P}_{j}
\begin{pmatrix} r_{b1} \\ r_{b2} \\ r_{b3} \\ r_{b4} \end{pmatrix}_{j}
\end{aligned}
\end{eqnarray}
or
\begin{eqnarray}
\begin{aligned}
	\begin{pmatrix} r_{b1} \\ r_{b2} \\ r_{b3} \\ r_{b4} \end{pmatrix}_{j}=\mathbf{P}_{j}^{-1}...\mathbf{P}_{n-1}^{-1}\mathbf{P}_{n}^{-1}\mathbf{L}_{n+1}
\begin{pmatrix} 0 & 0 \\ 0 & 0 \\ 1 & 0 \\ 0 & 1 \end{pmatrix}\begin{pmatrix} \grave{P} \\ \grave{S} \end{pmatrix} \cdot
\end{aligned}
\end{eqnarray}

The displacement-stress vector from the free surface to the source/receiver are linked by:

\begin{eqnarray}
\begin{aligned}
	\begin{pmatrix} r_{u1} \\ r_{u2} \\ r_{u3} \\ r_{u4} \end{pmatrix}_{j}=\mathbf{P}_{j}\mathbf{P}_{j-1}...\mathbf{P}_{1}
\begin{pmatrix} 1 & 0 \\ 0 & 1 \\ 0 & 0 \\ 0 & 0 \end{pmatrix}\begin{pmatrix} u \\ w \end{pmatrix} \cdot
\end{aligned}
\end{eqnarray}

The propagation of the fundamental independent solutions of the displacement-stress at the surface down to the source can be defined as the columns of:

\begin{eqnarray}
\begin{pmatrix} Y_{u11} & Y_{u12} \\ Y_{u21} & Y_{u22} \\ Y_{u31} & Y_{u32} \\ Y_{u41} & Y_{u42} \end{pmatrix}=
 \mathbf{P}_{j-1}\mathbf{P}_{j-2}...\mathbf{P}_{1}\begin{pmatrix} 1 & 0 \\ 0 & 1 \\ 0 & 0 \\ 0 & 0 \end{pmatrix}\cdot
\end{eqnarray}

Similarly, the motion displacement-stress just below the source compatible with unitary 
downgoing P and S waves at half-space are the columns of 

\begin{eqnarray}
	\begin{pmatrix} Y_{b11} & Y_{b12} \\ Y_{b21} & Y_{b22} \\ Y_{b31} & Y_{b32} \\ Y_{b41} & Y_{b42} \end{pmatrix} =\mathbf{P}_j^{-1}...\mathbf{P}_{n-1}^{-1}\mathbf{P}_n^{-1}\mathbf{L}_{n+1}\begin{pmatrix} 0 & 0 \\ 0 & 0 \\ 1 & 0 \\ 0 & 1 \end{pmatrix}
\end{eqnarray}

For a horizontal harmonic load,  the displacements are assumed to be continuous at 
the source. The solution above and below the source are respectively:

\begin{eqnarray}
\begin{pmatrix} r_{u1} \\ r_{u2} \\ r_{u3} \\ r_{u4} \end{pmatrix}_{z}=
\begin{pmatrix} g_{11\text{PSV}} \\ g_{31\text{PSV}}/i \\ \sigma_{uh} \\ 0  \end{pmatrix}
\end{eqnarray}

and 

\begin{eqnarray}
\begin{pmatrix} r_{b1} \\ r_{b2} \\ r_{b3} \\ r_{b4}  \end{pmatrix}_{z}=
\begin{pmatrix} g_{11\text{PSV}} \\ g_{31\text{PSV}}/i\\ \sigma_{bh} \\ 0  \end{pmatrix}
.
\end{eqnarray}

The continuity of the stresses leads to the following boundary conditions:
\begin{eqnarray}
\begin{aligned}
r_{u4} - r_{b4} &= 0\\
\sigma_{bh} - \sigma_{uh} &= r_{b3} - r_{u3} = 1\\ 
\end{aligned}
\end{eqnarray}

The first two equations above can be written as:
\begin{eqnarray}
\begin{aligned}
	\mathbf{Ax}=\mathbf{b}_h
\label{eq:axbh}
\end{aligned}
\end{eqnarray}

where 

\begin{eqnarray}
\begin{aligned}
	\mathbf{A}=(\mathbf{Y}_b,-\mathbf{Y}_u)=\begin{pmatrix} Y_{b11} & Y_{b12} & -Y_{u11} & -Y_{u12} \\ Y_{b21} & Y_{b22} & -Y_{u21} & -Y_{u22} \\Y_{b31} & Y_{b32} & -Y_{u31} & -Y_{u32} \\ Y_{b41} & Y_{b42} & -Y_{u41} & -Y_{u42} \end{pmatrix}\cdot
\end{aligned}
\end{eqnarray}

\begin{eqnarray}
\begin{aligned}
	\mathbf{x}=\begin{pmatrix} \grave{P} \\ \grave{S} \\ u \\ w \end{pmatrix}
\end{aligned}
\end{eqnarray}

and 

\begin{eqnarray}
\begin{aligned}
	\mathbf{b}_h=\begin{pmatrix} 0 \\ 1 \\ 0 \\ 0 \end{pmatrix}\cdot
\end{aligned}
\end{eqnarray}

Similarly, for a vertical harmonic load (upper layer at load) it follows that

\begin{eqnarray}
\begin{pmatrix} r_{u1} \\ r_{u2} \\ r_{u3} \\ r_{u4} \end{pmatrix}_z=
\begin{pmatrix} g_{13\text{PSV}} \\ g_{33\text{PSV}}/i \\ 0 \\ \sigma_{uv}/i  \end{pmatrix}
\end{eqnarray}

and 

\begin{eqnarray}
\begin{pmatrix} r_{b1} \\ r_{b2} \\ r_{b3} \\ r_{b4} \end{pmatrix}_z=
\begin{pmatrix} g_{13\text{PSV}} \\ g_{33\text{PSV}}/i \\ 0 \\ \sigma_{bv}/i  \end{pmatrix}\cdot
\end{eqnarray}

In this case, the boundary conditions are:
\begin{eqnarray}
\begin{aligned}
r_{u4} - r_{b4} &= 1\\
\sigma_{bv} - \sigma_{uv} &= r_{b3} - r_{u3} = 0\\ 
r_{u1} &= r_{b1} = g_{13\text{PSV}}\\
r_{u2} &= r_{b2} = g_{33\text{PSV}}
\end{aligned}
\end{eqnarray}

From the first two equations, it is possible to write, as for the horizontal load: 
\begin{eqnarray}
\begin{aligned}
	\mathbf{Ax}=\mathbf{b}_v
\label{eq:axbv}
\end{aligned}
\end{eqnarray}

Where $\mathbf{A}$ and $\mathbf{x}$ have been defined above. $\mathbf{b}_v$ is defined in this case by:

\begin{eqnarray}
\begin{aligned}
	\mathbf{b}_v=\begin{pmatrix} 1 \\ 0 \\ 0 \\ 0 \end{pmatrix}
\end{aligned}
\end{eqnarray}

Equations~\ref{eq:axbh} and \ref{eq:axbv} can be solved for $g_{11\text{PSV}}$ and $g_{33\text{PSV}}$ by using, for example, the Gaussian LU matrix decomposition.

The Green's function in 1D layered medium are then given by:
\begin{eqnarray}
 \text{Im}\left[G_{22}^{\text{P-SV}}(z, f)\right] = \text{Im}\left[G_{11}^{\text{P-SV}}(z, f)\right] = \frac{1}{4\pi}\int_{0}^{\infty} 
 \text{Im}\left[g_{11\text{PSV}}\right]kdk
\end{eqnarray}

\begin{eqnarray}
 \text{Im}\left[G_{33}^{\text{P-SV}}(z_F,f)\right] =  \frac{1}{2\pi}\int_{0}^{\infty} \text{Im}\left[g_{33\text{PSV}}\right]kdk
\end{eqnarray}

The solution to the integral can be obtained numerically by using for example the discrete wavenumber approach.

\section{Orthonormalization algorithm for the P-SV waves propagation}
\label{appen:orthonormalization}

\subsection{Propagation from the surface to the source}
Starting from the definition of the base vector $\mathbf{Y}_{j}$ at the layer interface $j$ (Appendix~\ref{appen:p-sv}),
\begin{eqnarray}
	\mathbf{Y}_{j+1} =  \mathbf{P}_j\mathbf{Y}_j = \mathbf{L}_j\mathbf{E}_j\mathbf{L}_j^{-1}\mathbf{Y}_j
\end{eqnarray}

Let define $\mathbf{C}_{j}$ such that:
\begin{eqnarray}
	\mathbf{C}_{j} = \mathbf{L}_j^{-1}\mathbf{Y}_{j} 
\end{eqnarray}
and
\begin{eqnarray}
	\mathbf{Y}_{j+1} = \mathbf{L}_j\mathbf{E}_j\mathbf{C}_{j}
\end{eqnarray}

Redefine $\mathbf{Y}$ to $\mathbf{Y}'$  so that

\begin{eqnarray}
	\mathbf{Y}_{j+1}' = \mathbf{L}_j\mathbf{E}_j\mathbf{C}_{j}'
\end{eqnarray}

Where $\mathbf{C}_{j}'$ is defined such that

\begin{eqnarray}
	\mathbf{C}_{j}' = \mathbf{C}_{j}\mathbf{Q_u} = \begin{pmatrix} C_{11} & C_{12} \\ C_{21} & C_{22} \\ C_{31} & C_{32} \\ C_{41} & C_{42} \end{pmatrix}\begin{pmatrix} Q_{u11} & Q_{u12} \\ Q_{u21} & Q_{u22}  \end{pmatrix}=\begin{pmatrix} 1 & 0 \\ 0 & 1 \\ C_{31}' & C_{32}' \\ C_{41}' & C_{42}' \end{pmatrix}
\end{eqnarray}

This equation leads to

\begin{eqnarray}
Q_{u11} = \frac{C_{22}}{C_{11}C_{22}-C_{12}C_{21}},\\
Q_{u12} = \frac{-C_{12}}{C_{11}C_{22}-C_{12}C_{21}},\\
Q_{u21} = \frac{-C_{21}}{C_{11}C_{22}-C_{12}C_{21}},\\
Q_{u22} = \frac{C_{11}}{C_{11}C_{22}-C_{12}C_{21}}
\end{eqnarray}

$\mathbf{C}_{j}'$ contains in each column different wave types together with the corresponding reflections.

\subsection{Propagation from the half-space to the source}
For the wave propagation from the half-space to the source, the matrix of basis vectors can be written as: 

\begin{eqnarray}
	\mathbf{Y}_{j} = \mathbf{L}_j\mathbf{E}_j^{-1}\mathbf{L}_j^{-1}\mathbf{Y}_{j+1}
\end{eqnarray}
In this case, $\mathbf{C}_{j}$  is defined such that:  
\begin{eqnarray}
	\mathbf{C}_{j} = \mathbf{L}_j^{-1}\mathbf{Y}_{j} 
\end{eqnarray}
This leads to
\begin{eqnarray}
	\mathbf{Y}_{j}=\mathbf{L}_j\mathbf{C}_{j}
\end{eqnarray}
For the layer $j+1$, we have:

\begin{eqnarray}
	\mathbf{Y}_{j+1}=\mathbf{L}_{j+1}\mathbf{C}_{j+1}
\end{eqnarray}
We then obtain:
\begin{eqnarray}
	\mathbf{Y}_{j} = \mathbf{L}_j\mathbf{E}_j^{-1}\mathbf{L}_j^{-1}\mathbf{L}_{j+1}^{-1}\mathbf{C}_{j+1}
\end{eqnarray}

Reset $\mathbf{C}_j$
\begin{eqnarray}
	\mathbf{C}_{j} = \mathbf{L}_j^{-1}\mathbf{L}_{j+1}\mathbf{C}_{j+1} 
\end{eqnarray}

So that

\begin{eqnarray}
	\mathbf{Y}_{j} = \mathbf{L}_j\mathbf{E}_j^{-1}\mathbf{C}_{j}
\end{eqnarray}

Redefine $\mathbf{Y}'$ so that

\begin{eqnarray}
	\mathbf{Y}_{j}' = \mathbf{L}_j\mathbf{E}_j^{-1}\mathbf{C}_{j}'
\end{eqnarray}

Where $\mathbf{C}_{j}'$ is defined such that

\begin{eqnarray}
	\mathbf{C}_{j}' = \mathbf{C}_{j}\mathbf{Q_b} = \begin{pmatrix} C_{11} & C_{12} \\ C_{21} & C_{22} \\ C_{31} & C_{32} \\ C_{41} & C_{42} \end{pmatrix}\begin{pmatrix} Q_{b11} & Q_{b12} \\ Q_{b21} & Q_{b22}  \end{pmatrix}=\begin{pmatrix} C_{11}' & C_{12}' \\ C_{21}' & C_{22}' \\ 1 & 0 \\ 0 & 1 \end{pmatrix}
\end{eqnarray}

This equation leads to

\begin{eqnarray}
Q_{b11} = \frac{C_{42}}{C_{31}C_{42}-C_{32}C_{41}},\\
Q_{b12} = \frac{-C_{32}}{C_{31}C_{42}-C_{32}C_{41}},\\
Q_{b21} = \frac{-C_{41}}{C_{31}C_{42}-C_{32}C_{41}},\\
Q_{b22} = \frac{C_{31}}{C_{31}C_{42}-C_{32}C_{41}}.
\end{eqnarray}

In this representation, $\mathbf{C}_{j}'$ contains in each column different wave types separately together with their corresponding reflections.

\section{Pseudo 4x4 propagator matrix for a water layer 
on top of a layered elastic medium}
\label{sect:pseudo}
In presence of a water layer, characterized by a shear 
stress $\mu = 0$, only $P-$waves contribute to the Green's function estimation. 

Starting from the wave equation for the P-SV case, it can be demonstrated  that:

\begin{eqnarray}
r_1 = \dfrac{k}{\rho\omega^2}r_4
\end{eqnarray}

and

\begin{eqnarray}
\begin{aligned}
\dfrac{\partial r_4}{\partial z} &= -\rho\omega^2r_2 \\
\dfrac{\partial r_2}{\partial z} &= \dfrac{1}{\rho\omega^2}\ \left( -k^2 + \dfrac{\omega^2}{\alpha^2} \ \right)r_4
\end{aligned}
\end{eqnarray}

\begin{eqnarray}
\dfrac{d}{d z} \dbinom{r_2}{r_4} = \begin{pmatrix} 0&\dfrac{1}{\rho\omega^2}\ \left( -k^2 + \dfrac{\omega^2}{\alpha^2} \ \right)\\
 - \omega^2\rho& 0\\\end{pmatrix} \dbinom{r_2}{r_4} 
\end{eqnarray}

This equation  is of the form:

\begin{eqnarray}
\dfrac{d \mathbf{r}}{d z} = \mathbf{A} \mathbf{r} 
\label{eq:eqdiffmatrix-water}
\end{eqnarray}
where 
\begin{eqnarray}
\mathbf{r} = \dbinom{r_2}{r_4}
\end{eqnarray}
and 
\begin{eqnarray}\mathbf{A}= \begin{pmatrix} 0&\dfrac{1}{\rho\omega^2}\ \left( -k^2 + \dfrac{\omega^2}{\alpha^2} \ \right)\\
- \omega^2\rho& 0\end{pmatrix}\cdot
\end{eqnarray}
The solution to Equation \ref{eq:eqdiffmatrix-water} at two points $z_1$ (at 
the water surface) and $z_2$ (at the ocean floor) is
\citep{bib:gantmacher1959,bib:gilbertandbackus1966,bib:akiandrichards2002}:
\begin{eqnarray}
\dbinom{r_2}{r_4}_{z_2} = \mathbf{P} \dbinom{r_2}{r_4}_{z_1} 
\label{eq:prop-water}
\end{eqnarray}
where
\begin{eqnarray}
	\mathbf{P}= \begin{pmatrix} \cosh[\gamma(z_2-z_1)]& -\dfrac{\gamma}{\rho\omega^2}\sinh[\gamma(z_2-z_1)]\\
-\dfrac{\rho\omega^2}{\gamma} \sinh[\gamma(z_2-z_1)] & \cosh[\gamma(z_2-z_1)] \end{pmatrix}
\end{eqnarray}

To obtain the pseudo 4x4 matrix, we rewrite Equation \ref{eq:prop-water} as 
follows (see also \citealt{bib:herrmann2008})

\begin{eqnarray}
\begin{pmatrix} r_1|_{z_{2}} \\ r_2|_{z_{2}} \\ r_3|_{z_{2}} \\ r_4|_{z_{2}} \end{pmatrix} = \begin{pmatrix} 1 & 0 & 0 & 0\\
					0 & \cosh[\gamma(z_2-z_1)] & 0 & -\dfrac{\gamma}{\rho\omega^2}\sinh[\gamma(z_2-z_1)]\\
					0 & 0 & 1 & 0 \\
					0 & -\dfrac{\rho\omega^2}{\gamma} \sinh[\gamma(z_2-z_1)] & 0 & \cosh[\gamma(z_2-z_1)]\end{pmatrix} 
	\begin{pmatrix} r_1|_{z_{2}} \\ r_2|_{z_{1}} \\ r_3|_{z_{2}} \\ r_4|_{z_{1}}\end{pmatrix}
\end{eqnarray}

The pseudo-propagator matrix in terms of eigenvector ($\mathbf{L}$) and eingenvalues ($\mathbf{E}$)  matrices can be given by:

\begin{eqnarray}
	\begin{aligned}
		\mathbf{P}_{\text{pseudo}} &= \begin{pmatrix} 1 & 0 & 0 & 0\\
          0 & \cosh[\gamma(z_2-z_1)] & 0 & -\dfrac{\gamma}{\rho\omega}\sinh[\gamma(z_2-z_1)]\\
          0 & 0 & 1 & 0 \\
          0 & -\dfrac{\rho\omega^2}{\gamma} \sinh[\gamma(z_2-z_1)] & 0 & \cosh[\gamma(z_2-z_1)]\end{pmatrix}	\\
							&=\mathbf{LEL}^{-1}\\
							&=\begin{pmatrix} 1 & 0 & 0 & 0\\
          0 & -\dfrac{\gamma}{\rho\omega^2} & 0 & \dfrac{\gamma}{\rho\omega^2}\\
          0 & 0 & 1 & 0 \\
          0 & 1 & 0 & 1\end{pmatrix}\\
		&\ \begin{pmatrix} 1 & 0 & 0 & 0\\
          0 & \exp[\gamma(z_2-z_1)] & 0 & 0\\
          0 & 0 & 1 & 0 \\
          0 & 0 & 0 & \exp[-\gamma(z_2-z_1)]\end{pmatrix}\\
	&\ \begin{pmatrix} 1 & 0 & 0 & 0\\
          0 & -\dfrac{\rho\omega^2}{2\gamma} & 0 & \dfrac{1}{2}\\
          0 & 0 & 1 & 0 \\
          0 & \dfrac{\rho\omega^2}{2\gamma} & 0 & \dfrac{1}{2}\end{pmatrix}
	\end{aligned}
\end{eqnarray}
Where $\gamma=\sqrt{k^2-\omega^2/V_\mathbf{P}^2}$.

From this point on, the algebra is again similar to the derivations presented earlier. The effect of the presence of the 
water layer on the estimated $H/V$ spectral ratio curves is discussed in the text.


\bsp 
\label{lastpage}
\end{document}